\documentclass{article}

    \PassOptionsToPackage{numbers, compress}{natbib}

    \usepackage[preprint]{neurips_2024}

\usepackage[utf8]{inputenc} 
\usepackage[T1]{fontenc}    
\usepackage{hyperref}       
\usepackage{url}            
\usepackage{booktabs}       
\usepackage{amsfonts}       
\usepackage{nicefrac}       
\usepackage{microtype}      
\usepackage{xcolor}         

\usepackage{float}  		%
\usepackage{nicematrix}  	
\usepackage{amsmath}
\usepackage{graphicx}
\usepackage{rotating}
\usepackage{multirow}
\usepackage{lscape}
\usepackage{subcaption}
\usepackage[ruled, vlined, linesnumbered]{algorithm2e}	
\usepackage{amsthm}		
\newtheorem{theorem}{Theorem}
\newtheorem{lemma}{Lemma}
\usepackage{placeins}
\usepackage{makecell}

\newcommand{\bigO}[1]{\mathcal{O}(#1)}
\newcommand{\gradj}[1]{A_{G_{#1}}}
\newcommand{\gdag}[0]{\overrightarrow{H}}
\newcommand{\gdagve}[0]{\overrightarrow{H}=\{V,\overrightarrow{E}\}}
\newcommand{\ed}[0]{\overrightarrow{E}}
\newcommand{\edag}[0]{\overrightarrow{E}}

\newcommand{\xyz}[0]{x, y \text{ and } z}
\newcommand{\orbit}[1]{o_{#1}}
\newcommand{\adjSingle}[2]{A_{o_{#1 \scalebox{.8}{-} #2}}}
\newcommand{\adjDouble}[2]{A_{o_{#1 \scalebox{.8}{-\,-} #2}}}
\newcommand{\adjTriple}[2]{A_{o_{#1 \scalebox{.8}{-\,-\,-} #2}}}
\newcommand{\gitertwo}{\sum\limits_{\substack{x,z: y<z, \\ G[\{x,y,z\}] \\ \cong G_2}}}
\newcommand{\giteroneA}{\sum\limits_{\substack{x,z: y<z, \\ z \in N(x), \\ G[\{x,y,z\}] \cong G_1}}}
\newcommand{\giteroneB}{\sum\limits_{\substack{x,z: y<z, \\ z \in N(y), \\ G[\{x,y,z\}] \cong G_1}}}
\newcommand{\rowsum}{\sum\limits_{y=1}^{n}}
\DeclareCaptionType{equationset}[Equation Set]

\newcommand*\fref[1]{Fig. \ref{#1}}   

\title{Graphlets correct for the topological information missed by random walks}

\author{%
  Sam F. L. Windels \\
  Life Sciences\\
  Barcelona Supercomputing Center\\
  08034 Barcelona, Spain\\
  \texttt{sam.windels@bsc.es} \\
  \And
  No{\"e}l Malod-Dognin \\
  Life Sciences\\
  Barcelona Supercomputing Center\\
  08034 Barcelona, Spain\\
  \texttt{noel.malod@bsc.es} \\
  \AND
  Nata\v{s}a Pr\v{z}ulj\\
  Barcelona Supercomputing Center\\
  08034 Barcelona, Spain\\
  Department of Computer Science\\
  University College London\\
  WC1E 6BT London, UK \\
  ICREA, Pg Llu\'is Companys 23\\
  08010 Barcelona, Spain\\
  \texttt{natasha@bsc.es} \\
}

\begin{document}

\maketitle

\begin{abstract}

Random walks are widely used for mining networks due to the computational
efficiency of computing them. For instance, graph representation learning learns a
$d$-dimensional embedding space, so that the nodes that tend to co-occur on
random walks (a proxy of being in the same network neighborhood) are close in
the embedding space. Specific local network topology (i.e., structure)
influences the co-occurrence of nodes on random walks, so random walks of limited length capture
only partial topological information, hence diminishing the performance of
downstream methods. We explicitly capture all topological neighborhood
information and improve performance by introducing orbit adjacencies that quantify
the adjacencies of two nodes as co-occurring on a given pair of \emph{graphlet
orbits}, which are symmetric positions on \emph{graphlets} (small, connected,
non-isomorphic, induced subgraphs of a large network). Importantly, we
mathematically prove that random walks on up to $k$ nodes capture only a subset
of all the possible orbit adjacencies for up to $k$-node graphlets. Furthermore,
we enable orbit adjacency-based analysis of networks by developing an efficient
GRaphlet-orbit ADjacency COunter (GRADCO), which exhaustively computes all 28
orbit adjacency matrices for up to four-node graphlets. Note that four-node
graphlets suffice, because real networks are usually small-world. In large
networks on around 20,000 nodes, GRADCO computes the 28 matrices in minutes. 
On six real networks from various domains, we compare the performance of
node-label predictors obtained by using the network embeddings based on our
orbit adjacencies to those based on random walks. We find that orbit
adjacencies, which include those unseen by random walks, outperform random
walk-based adjacencies, demonstrating the importance of the inclusion of the
topological neighborhood information that is unseen by random walks. Our
results open up multiple research questions, e.g.: (a) the inclusion of orbit
adjacency information in more scalable, sampling-based architectures; (b) how
to best jointly mine different types of topological neighborhoods; and (c) how
to select the specific topological neighborhood \emph{a priori} that works best for a
given task.

\end{abstract}

\section{Introduction}

Networks are a natural way to model complex systems across a wide range of
domains, such as social science \cite{tabassum2018social}, biology \cite{johnson2024graph} and epidemiology \cite{liu2024review}.
\emph{Random walks}, the stochastic process that selects a random starting node and
repeatedly randomly traverses edges to neighbouring nodes, are a simple and
efficient way to capture network neighbourhood structure, i.e., what nodes
cluster in a network. Intuitively, as nodes that cluster in the network are
more likely to be connected to nodes within the cluster than to nodes outside
the cluster, nodes within the same cluster should tend to co-occur on random
walks \cite{lovasz1993random}. As such, random walks are used in a wide range
of network analysis tasks, such as node attribute prediction, link prediction
and community detection \cite{xia2019random}.

Because of the computational complexity of mining a network directly, current state-of-the-art methods for analyzing networks are based on graph representation learning (also called network embedding) \cite{Nelson_2019_to_embed}.
The random-walk based graph representation learning methods take
inspiration from Natural Language Processing (NLP) methods. In NLP, neural
architectures are used to learn an embedding space so that words that tend to
co-occur in sentences are close in the embedding space. In practice, word
sequences of a given length are sampled from a text corpus and the neural
architecture, e.g., the skip-gram model, is trained to predict the context words
that surround the target word at the centre of the sampled sequence
\cite{rong2014word2vec}. Analogously, random walk based node embedding methods
sample random walks from a network and learn to predict what nodes will occur
on the walk given the starting node. For instance, DeepWalk uses the skip-gram
architecture to learn an embedding space, so that nodes that tend to co-occur on
random walks up to a given length, $l$, are close in the embedding space
\cite{perozzi2014deepwalk}. By changing the way the random walk algorithm traverses the
network, i.e., by changing the random walk strategy, different notions of what it means for 
two nodes to be in each other's neighbourhood can be captured. 
Node2vec, for instance, allows steering the random walks to stay close, or to
stray further away into the network with the tuning of two hyperparameters,
allowing to imprecisely capture a more local or a more global node neighbourhood \cite{Grover_2016_node2vec}.

Another approach to mine networks is based on
precisely capturing the entire local structure (topology, wiring) of a network. 
The state-of-the-art methods to precisely capture network
wiring are based on \emph{graphlets}: small, connected, non-isomorphic, induced sub-graphs
of a large network \citep{Przulj_2004_modeling}. To increase the level of
detail captured, the state-of-the-art methods distinguish symmetric node
positions within each graphlet, known as \emph{automorphism orbits}
\citep{prvzulj2007biological}, that describe the different topological roles of
a node in a particular graphlet. A widely used node-level descriptor
summarising the wiring of a node in a network is the Graphlet Degree Vector
(GDV), which lists for a given node the frequency of it occurring on each of
the different graphlet orbits in a network \citep{milenkovic2008uncovering}. We illustrate the
concepts of graphlets, node orbits and the GDV vector in
\fref{fig:concepts}A-C. Analogous to node orbits, one can also consider
symmetric positions of edges within graphlets, known as \emph{edge orbits}
\citep{Solava_2012_edge_orbit}. For the remainder of the paper, we use the term
`orbits' to refer to node orbits. 

To precisely capture graphlet-based, extended 
neighbourhood information, Windels \emph{et al.} introduced \emph{graphlet adjacency}, which
considers two nodes adjacent based on their frequency of co-occurrence on a
given graphlet (illustrated in \fref{fig:concepts}D) \citep{Windels_2019_graphlet}. They used graphlet adjacency to define 
graphlet based spectral embeddings, which embed nodes close in an embedding
space if the nodes tend to co-occur on a given graphlet.

\subsection{Problem}

Compared to the clear structural interpretation of graphlet adjacency, random
walks only have a vague structural interpretation, e.g., capturing `local' versus
`global' network organisation depending on the length of the walks and how they
are steered. Moreover, given that the co-occurrence of nodes on random walks of limited length is
influenced by the local network topology, random walks implicitly capture only partial
topological information. 
Furthermore, the extent to which random walks capture topological information is
not well understood.

\subsection{Contribution}
\label{sec:contribution}

In this paper, we generalise graphlet adjacency to \emph{orbit adjacency}, which
quantifies the adjacencies of two nodes as their co-occurrence frequency on a given pair of
graphlet orbits (detailed below). We use orbit adjacency to explicitly define the topological
neighbourhood information captured by random walks and show that random walks
that cover up to \(k\) nodes capture only a subset of all possible orbit
adjacencies for up to \(k\)-node graphlets. To show the real-world value of 
orbit adjacencies, we define orbit adjacency-based network embeddings and show
that they outperform random walk-based embeddings in a multi-class node label
prediction task on six real networks from various domains. 

To enable orbit adjacency-based analysis of networks, we develop efficient
GRaphlet-orbit ADjacency COunter (GRADCO), which exhaustively computes all 28
orbit adjacency matrices for up to four-node graphlets in a matter of minutes on
large real and synthetic networks of around 20,000 nodes (see Section \ref{sub:gradco} in appendix for our algorithms and performance benchmarks). Our software is
implemented in C++ and available as a Python package on PyPI under the CC BY-NC-SA 4.0 licence.
As orbit adjacency generalises all existing graphlet-based counts, i.e., orbit
counts \cite{prvzulj2007biological}, edge orbit counts
\cite{Solava_2012_edge_orbit}, graphlet adjacency \cite{Windels_2019_graphlet}
and edge orbit adjacency \cite{Rossi_2020_structural}, our counter is
universally applicable to all graphlet-based network analysis tasks.

\begin{figure} \begin{tabular}{m{0.4cm} >{\centering\arraybackslash} m{\columnwidth}} 
	A) & 
	\raisebox{1.cm}{\(H\):}
	\includegraphics[width=.150\linewidth]{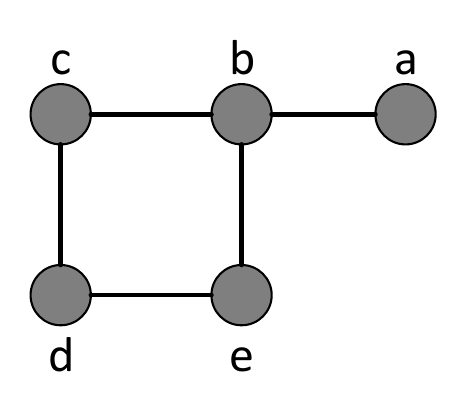} \\ 
	B) &
	\includegraphics[width=0.4\linewidth]{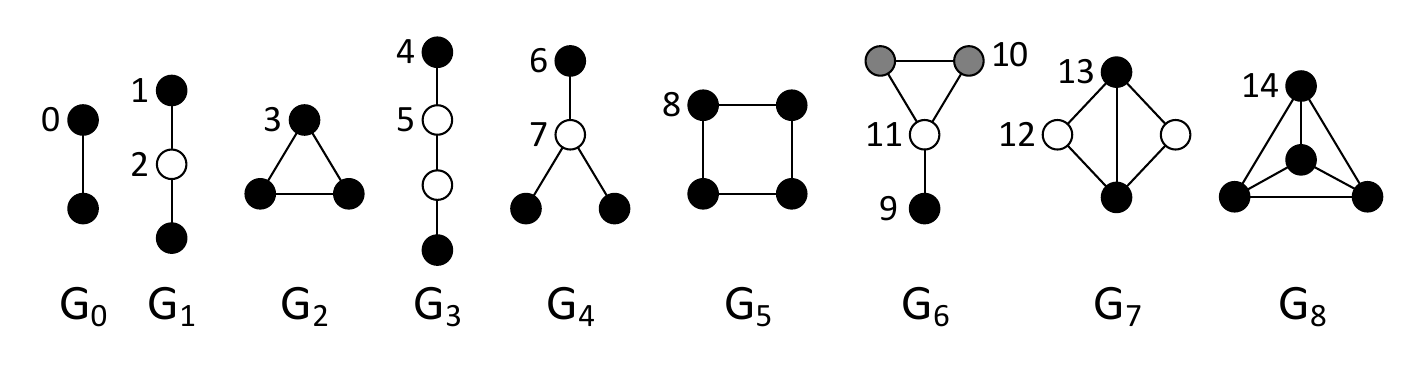} \\
	C) & 
	\includegraphics[width=0.4\linewidth]{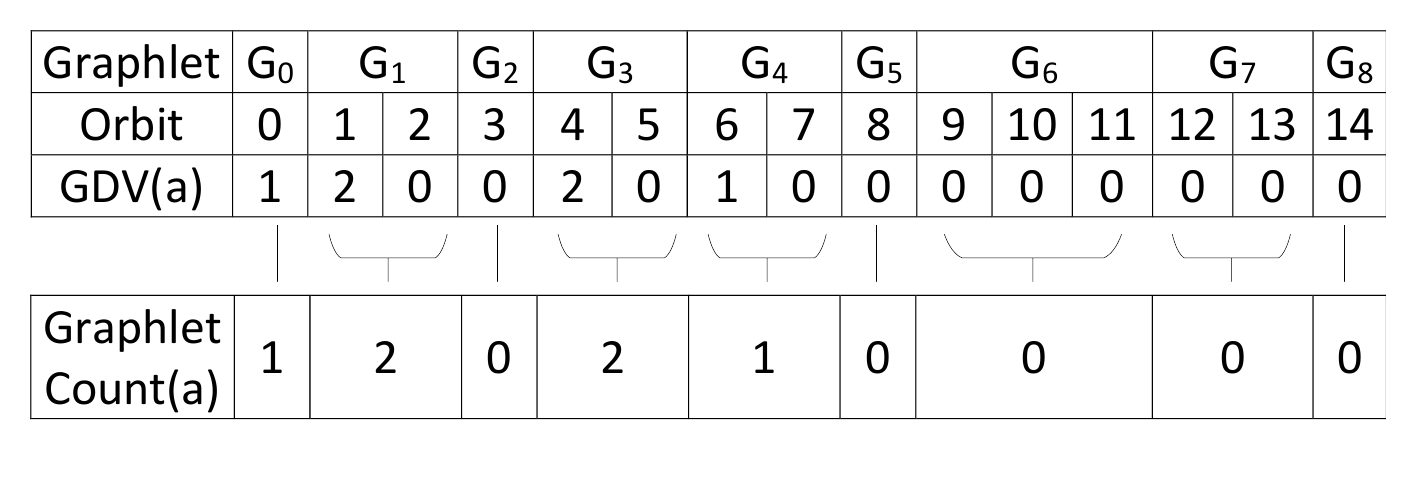} \\
	 D) &
	 \begin{tabular}{ r c} 
		\setlength{\arraycolsep}{2pt}
		\begin{tabular}{r}
			\( A = \)\\
			\(\gradj{0} = \) \\ 
			\(\adjSingle{0}{0} = \)
		\end{tabular}
		\(
		\begin{bNiceMatrix}[first-row,last-col]
			 a & b & c & d & e & \\
			 0 & 1 & 0 & 0 & 0 & a\\ 
			 1 & 0 & 1 & 0 & 1 & b\\ 
			 0 & 1 & 0 & 1 & 0 & c\\ 
			 0 & 0 & 1 & 0 & 1 & d\\ 
			 0 & 1 & 0 & 1 & 0 & e \\
		\end{bNiceMatrix}
		 \)
			 & 
			 \setlength\arraycolsep{2pt} 
			 \( \hspace{8pt} \gradj{1} =
		\begin{bNiceMatrix}[first-row,last-col]
			 a & b & c & d & e & \\
			 0 & 2 & 1 & 0 & 1 & a\\
			 2 & 0 & 3 & 2 & 3 & b\\ 
			 1 & 3 & 0 & 2 & 2 & c\\ 
			 0 & 2 & 2 & 0 & 2 & d\\ 
			 1 & 3 & 2 & 2 & 0 & e \\
		\end{bNiceMatrix}
		 \)
	 \end{tabular} \\
	 \vspace{\baselineskip}\\
	 E) &
	 \begin{tabular}{ r c} 
		\setlength{\arraycolsep}{2pt}
		\begin{tabular}{r}
			\(\adjSingle{1}{2} = \)
		\end{tabular}
		\(
		\begin{bNiceMatrix}[first-row,last-col]
			 a & b & c & d & e & \\
			 0 & 2 & 0 & 0 & 0 & a\\ 
			 0 & 0 & 1 & 0 & 1 & b\\ 
			 0 & 2 & 0 & 1 & 0 & c\\ 
			 0 & 0 & 1 & 0 & 1 & d\\ 
			 0 & 2 & 0 & 1 & 0 & e \\
		\end{bNiceMatrix}
		 \)
			 & 
			 \setlength\arraycolsep{2pt} \(\adjDouble{1}{1}=
		\begin{bNiceMatrix}[first-row,last-col]
			 a & b & c & d & e & \\
			 0 & 0 & 1 & 0 & 1 & a\\
			 0 & 0 & 0 & 2 & 0 & b\\ 
			 1 & 0 & 0 & 0 & 2 & c\\ 
			 0 & 2 & 0 & 0 & 0 & d\\ 
			 1 & 0 & 2 & 0 & 0 & e \\
		\end{bNiceMatrix}
		 \)
 	 \end{tabular} \\
 	 \end{tabular} 
	 \caption[An illustration of graphlets and orbit adjacencies.]{ 
		 { \bf An illustration of graphlets, orbits, graphlet adjacency and orbit adjacency.} 	 
	 {\bf A):} Example network \(H\). { \bf B): } All the graphlets with up to four-nodes, labelled
	 from \(G_0\) to \(G_8\). The automorphism orbits are indicated by the same shade and labelled from 0 to 14.
	 {\bf C): } The frequency at which node \(a\) occurs on each orbit in the example
	 network H (panel A) and how those counts sum into
 graphlet counts.
	 For instance, node \(a\) occurs on orbit \(\orbit{2}\) twice: once in the
	 path \(a\)-\(b\)-\(c\) and once in the path \(a\)-\(b\)-\(e\). It never occurs on
	 orbit \(\orbit{1}\), i.e., the centre of a three-node path. Hence, node
	 \(a\) occurs on graphlet \(G_1\) twice. {\bf D): } The graphlet adjacency
	 matrices \(\gradj{0}\) and \(\gradj{1}\) for the example network
	 \(H\). The off-diagonal elements of \(\gradj{1}\) correspond to the
	 frequency at which the nodes in the corresponding rows and columns co-occur on
	 graphlet \(G_1\) in \(H\). For instance, \(\gradj{1}(a,b)=2\), as
	 \(a\) and \(b\) co-occur twice on \(G_1\): via paths \(a\)-\(b\)-\(c\)
	 and \(a\)-\(b\)-\(e\). {\bf E):} The orbit adjacency matrices
	 \(\adjSingle{1}{2}\) and \(\adjDouble{1}{1}\) for the example network \(H\).
	 {\bf} The off-diagonal elements of \(\adjSingle{1}{2}\) correspond to
	 the frequency at which the nodes in the corresponding rows and columns co-occur
	 on graphlet \(G_1\) in \(H\), with the \(row\)-node touching orbit 1 and the
	 \(column\)-node touching orbit 2. For instance,
	 \(\adjSingle{1}{2}(a,b)=2\), as \(a\) and \(b\) co-occur twice on \(G_1\) with
	 node \(a\) on orbit 1 and node \(b\) on orbit 2: via paths \(a\)-\(b\)-\(c\) and
	 \(a\)-\(b\)-\(e\). Analogously, \(\adjDouble{1}{1}(a,b)=0\), as \(a\) and \(b\)
	 never co-occur on \(G_1\) with both \(a\) and \(b\) on orbit 1. Similar to
	 how orbit counts sum into graphlet counts (panel C), graphlet
	 adjacency matrix \(\gradj{1}\) can be computed as the sum of orbit
	 adjacency matrices: \(\gradj{1} = \adjSingle{1}{2} + \adjSingle{2}{1}
	 + \adjDouble{1}{1}\).
	} 
\label{fig:concepts} \end{figure} 

\section{Orbit adjacency definition}
\label{sec:orbit_adjacency}

To better take the topological roles of nodes into account, we generalise
graphlet adjacency to orbit adjacency.  Formally, given two nodes $u$ and $v$
in network $H$ and given two orbits $\orbit{i}$ and $\orbit{j}$ in graphlet
$G_k$, we define the orbit adjacency matrix, $\adjSingle{i}{j}$, as the $n
\times n$-dimensional matrix, in which each off-diagonal element,
$\adjSingle{i}{j}(u, v)$, is equal to the frequency nodes $u$ and $v$ co-occur
on orbits $\orbit{i}$ and $\orbit{j}$ in graphlet $G_k$ in $H$. As a way of
notation, we write $\adjSingle{i}{j}$ or $\adjDouble{i}{j}$ to indicate if the
two orbits are one or two hops away from each other in $G_k$,
respectively. We illustrate orbit adjacency in \fref{fig:concepts}E.  Note that
by definition, orbit adjacency $\adjSingle{0}{0}$ is equal to the graphlet
adjacency $A_{G_0}$, which is also equal to the traditional adjacency matrix.
Our new orbit adjacency matrices generalise both the GDV vectors and the
graphlet adjacency matrices, since both can be computed from the orbit
adjacency matrices (see Sections \ref{sub:orbit_counts} and
\ref{sub:graphlet_adjacency}, respectively).

\section{Random walks capture limited topological neighbourhood information}
\label{randomwalkneighborhood}

Here, we use our new orbit adjacencies to explicitly define the topological neighbourhood information captured by random walks, and show that random walks of length \(l\) that cover up to \(k=l+1\) distinct nodes capture only a subset of all the possible orbit adjacencies for up to \(k\)-node graphlets. 

First, we argue that random walk based methods, including the state-of-the-art network embedding methods, only consider information from the source and sink nodes from random walks. To show this, we start by stating the following well-known lemma:
\begin{lemma}
Let \(A\) be the binary adjacency matrix of an unweighted, undirected graph
\(H\). Then for the \(l^{th}\) power of the adjacency matrix, \(A^l\), the
entry \(A^l(i,j)\) is the number of possible unique walks of length \(l\) from
node \(i\) to node \(j\) in \(H\). 
\label{lem:random_walks}
\end{lemma}

We  refer to  \(A^l\) as the \emph{random walk adjacency matrix} for walks of length \(l\) in the remainder of this paper. Importantly, \(A^l\) only capture topological information from the source and sink nodes of the random walks (i.e., the number of unique walks of length \(l\) in between them).

Importantly, most of the random walk based approaches exploit a power of the row-normalised
random walk adjacency matrix, known as the \emph{transition matrix}, \(T =
D^{-1}A\). For example, as we further detail in section \ref{sec:embeddings},
DeepWalk learns node embeddings for walks up to length $l$ by factorising the
sum of the powers of the transition matrix \(T\) up to \(l\), i.e.,
\(\sum_{i=1}^{l} T^i\). The purpose of normalising the adjacency is to take
into account the scale-freeness of the network, steering the random walks to be
more likely to visit low-degree nodes than they would without the
normalization. However, this normalization does not allow for capturing
additional topological information than the non-normalized random walk
adjacency matrix, and hence, all methods exploiting random walk adjacency
matrices or transition matrices only consider the topological information from
the source and sink nodes of the random walks.

\begin{figure}[htpb]
	\centering
	\includegraphics[width=1.0\textwidth]{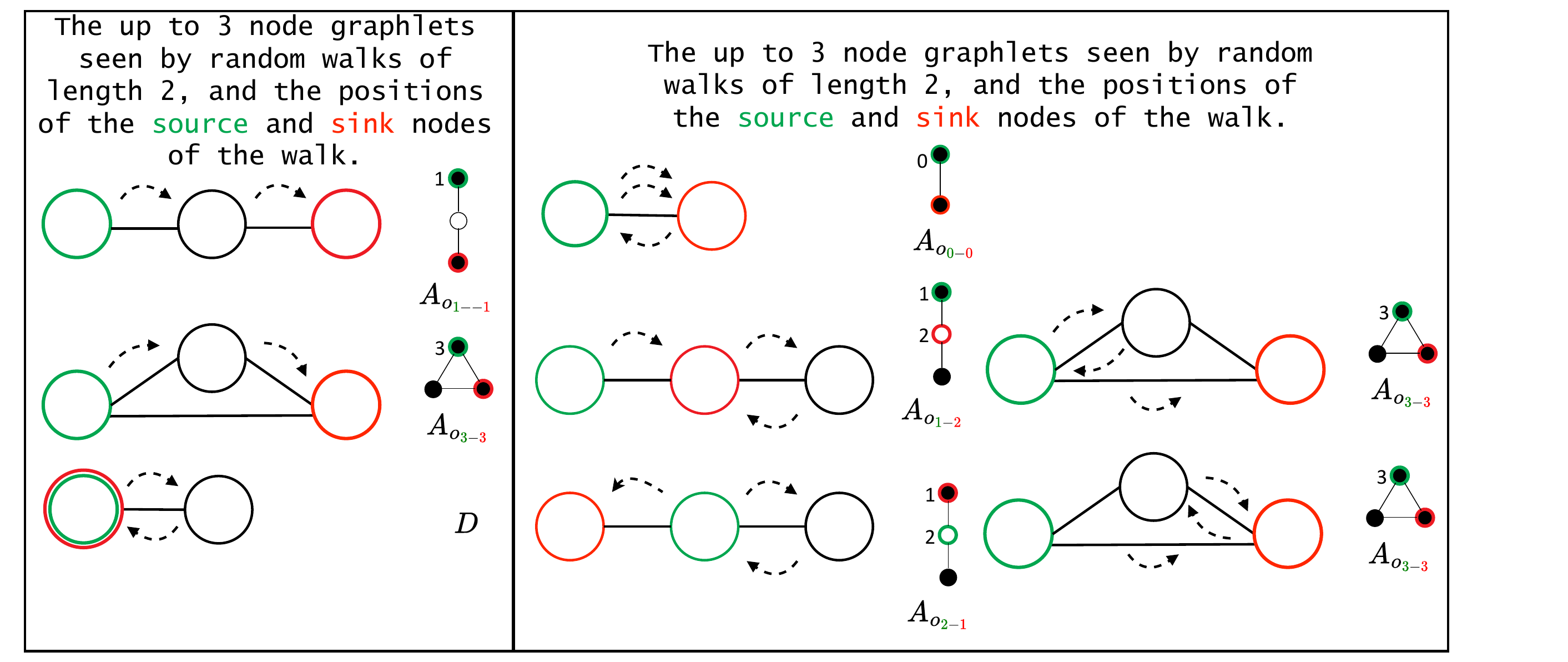}
	\caption{The different orbit adjacencies touched by the source and sink nodes in random walks of length 2 and 3. The dashed arrows indicate the edges visited by the random walks.}
	\label{fig:A2_walks}
\end{figure}

Second, we use our orbit adjacency to characterise the topological information
captured by the source and sink nodes of a random walk. We recall that a random
walk of length \(l\) covers up to \(k=l+1\) distinct nodes, which is the case
when every node in the walk is visited only once. Depending on how the covered
nodes are connected in the network, they induce one of the up to \(k\)-node
graphlets on the network. In Fig. \ref{fig:A2_walks}, we show all possible
random walk sequences of length two and three and the orbit adjacencies their
source nodes and sink nodes touch. One exception is the walks of length two
that return back to the source, which do not correspond to any orbit
adjacency. The first observation is that random walks of length two suffice to
uncover source and sink nodes co-occurring on orbit adjacencies
\(\adjDouble{1}{1}\) and \(\adjSingle{1}{2}\). However, if a random walk
happens to start at orbit 2 (the center of a three-node path), it needs three
hops to fully cover the path and to end up at orbit 1. So 
for a random walk to uncover that the source and sink node are
orbit adjacent in terms of \(\adjSingle{2}{1}\), walks of length three are
required. A similar reasoning holds for orbit adjacencies \(\adjSingle{0}{0}\)
and \(\adjSingle{1}{2}\). This is problematic because, as we will demonstrate
below, many random walk based methods, including those for network embedding,
exploit only the information between the source and sink nodes of random walks, and
thus, they do not capture \(\adjSingle{2}{1}\), \(\adjSingle{1}{2}\) and
\(\adjSingle{0}{0}\) when considering random walks of length two.

Next, we explicitly characterize the orbit adjacency based topological neighbourhood information captured by the source and sink nodes of random walks. For the random walk adjacency matrix \(A^2\), we prove the following theorem.
\begin{theorem}
\label{thm:random_walks}
The random walk adjacency matrix \(A^2\) is as a linear combination
of orbit adjacencies \(\adjDouble{1}{1}\) and \(\adjSingle{3}{3}\) and of the diagonal degree matrix \(D\):
\begin{equation}A^2 = \adjDouble{1}{1} + \adjSingle{3}{3} + D. \end{equation}
\end{theorem}

\begin{proof}
By applying lemma \ref{lem:random_walks}, the entry \(A^2(i, j)\) is the
number of unique walks of length two from node \(i\) to node \(j\) in
the graph \(H\). As shown in Fig. \ref{fig:A2_walks}, if \(i \neq j\),
there are two topologically distinct ways for a walk of length two to
start at node \(i\) and end at node \(j\). In either case, the walk traverses
from node \(i\) to node \(j\) via a shared neighbour. If nodes \(i\) and \(j\) 
are connected, the nodes in the walk induce a triangle on \(H\), and nodes 
\(i\) and \(j\) are orbit adjacent in terms of \(\adjSingle{3}{3}\). 
If nodes \(i\) and \(j\) are not connected, the nodes in the walk induce a
three-node path on \(H\), and nodes \(i\) and \(j\) are orbit adjacent in
terms of \(\adjDouble{1}{1}\). In case \(i = j\), the walk traverses to a
neighbour of \(i\) and goes back. The number of unique walks of length two that 
start and end at the same node \(i\) is equal to the degree of node \(i\).
So, given that by definition \(\adjDouble{1}{1}(i,j)\) is the 
total number of unique three node paths that start at node \(i\) and end at
node \(j\) and \(\adjSingle{3}{3}(i,j)\) is the total number of unique triangles on which node \(i\) and
node \(j\) co-occur and \(D\) is the diagonal degree matrix, it follows that:
\(A^2 = \adjDouble{1}{1} + \adjSingle{3}{3} + D.\)
\end{proof}

Using the same logic, it is straightforward to show the following theorem.
\begin{theorem}
The random walk adjacency matrix \(A^3\) is a linear combination
of orbit adjacency matrices of graphlets with up to four nodes:
\begin{equation}
\begin{split}
A^3 & =
\adjSingle{0}{0} 
+ \adjSingle{1}{2}
+ \adjSingle{2}{1}
+ 2 \adjSingle{3}{3} 
+ \adjTriple{4}{4}
+ \adjSingle{8}{8} \\
& \phantom{ = } + 2 \adjDouble{9}{10}
+ 2 \adjDouble{10}{9}
+ \adjDouble{12}{12} 
+ \adjSingle{12}{13}
+ \adjSingle{13}{12}
+ \adjSingle{14}{14}
\end{split}
\end{equation}
\label{thm:random_walks_3}
\label{eq:random_walks_3}
\end{theorem}

From theorem \ref{thm:random_walks}, it is clear that \(A^2\) captures orbit adjacencies
\(\adjDouble{1}{1}\) and \(\adjSingle{3}{3}\), but misses the remaining 
orbit adjacencies of graphlets with up to three nodes, \(\adjSingle{0}{0}, \adjSingle{1}{2} \text{ and }
\adjSingle{2}{1}\). From theorem \ref{thm:random_walks_3}, it is clear that
\(A^3\) captures 12 out of the 28 possible orbit
adjacencies of graphlets with up to four nodes (some of them twice), 
but misses the remaining 16. We visualise the orbit adjacencies
captured and missed by random walks of length two and three in Fig.
\ref{fig:blind_adjacencies} of the Appendix. 

Hence, as per theorem \ref{thm:random_walks}, random walk based methods
using walks of length two capture orbit adjacencies
\(\adjDouble{1}{1}\) and \(\adjSingle{3}{3}\), but miss the
remaining orbit adjacencies for graphlets with up to three nodes, \(\adjSingle{0}{0},
\adjSingle{1}{2} \text{ and } \adjSingle{2}{1}\). Indeed, to capture
these unseen orbit adjacencies, random walks of length three would have to
be considered. However, as shown in theorem \ref{thm:random_walks_3},
random walks of length three in turn miss many of the four node orbit
adjacencies, again missing potentially relevant topological
information. By increasing the length of the random
walks, the number of different topological signals captured by them increases: 
while random walks of length two only capture two types of orbit adjacencies and miss three, random
walks of length three already capture twelve (several of them twice), but miss 16. 
In addition, the topological information
that random walks do manage to capture is
combined (amalgamated, mixed) in non-understandable ways, 
with some of the same information captured multiple times.
Also, they are stochastic, so in a real case scenario there are no guarantees about what they will actually
capture (out of the topology that they can capture).
Note that many of the orbit adjacencies captured by random walks might not
be relevant for the task at hand (detailed below).

In conclusion, regardless of the length of the random walks, random
walk based methods do not capture all possible topological
neighbourhood information in the network. Additionally, the longer the
walk, the more the different topological signals are mixed, potentially
blurring the relevant topological information for the task at hand.

\section{Network embedding}
\label{sec:embeddings}

To evaluate the importance of the topological information missed by random
walks, we introduce orbit adjacency based node embeddings, building upon the
state-of-the-art node embeddings methods. We briefly formally
define the state-of-the-art node embedding methods and then extend them to
introduce our orbit adjacency based node embeddings.

The state-of-the-art random walk based embedding methods learn an embedding space so that
nodes that tend to co-occur on random walks are close in the embedding space.
In practice, random walk based node embedding methods sample random walks from
a network and learn to predict what nodes will occur on the walk given the
starting node. It has been shown that such embedding methods
implicitly factorise a matrix whose entries are the Pointwise Mutual
Information (PMI) between the co-occurrences of nodes, shifted by a
global constant \cite{Levy_2014, qiu_2018_network}. PMI measures the strength of
association between a pair of discrete outcomes $x$ and $y$, e.g., the co-occurrence of
nodes $x$ and $y$ on a random walk, by comparing the log-likelihood of the observed
co-occurrence of the outcomes to the log-likelihood of the outcomes occurring
independently: \(pmi(x,y) = log\left(\frac{p(x,y)}{p(x)p(y)}\right) \) \cite{church1990word}.
In the case of LINE \cite{tang2015line}, which can be seen as
a random walk embedding method for walks of length one, i.e., capturing
the co-occurrences of nodes on edges, the PMI between a pair of nodes is computed as:
\begin{equation}
	 PMI(i,j) = log \left(  \frac{ vol(A) \times A_{ij}}{d_i, d_j} \right) - log(b),
\end{equation}
where \(A\) is the adjacency matrix of the network, \(vol(A)\) the sum of all
the entries of the adjacency matrix, \(d_i\) is the degree of node \(i\), and
\(b\) is a global constant.  
The closed form solution for the PMI matrix implicitly factorised by LINE is:
\begin{equation}
	LINE = log \left( vol(A)  D^{-1} A D^{-1} \right) - log(b),
\end{equation}
where \(D\) is the degree matrix of the network.
DeepWalk \cite{perozzi2014deepwalk} can be seen as an extension of LINE that captures the co-occurrence of nodes on walks up to length \(T\).
The closed form solution for the PMI matrix implicitly factorised by DeepWalk is:
\begin{equation}
	DeepWalk = log \left( vol(A) \left( \frac{1}{T} \sum\limits_{r=1}^{T} (D^{-1}A)^r \right) D^{-1} \right) - log(b),
\end{equation}

Finally, Qiu \emph{et al.} apply Singular Value Decomposition (SVD) to explicitly factorise
the non-negative entries of these PMI matrices to compute the node embeddings
equivalent to LINE and DeepWalk \cite{qiu_2018_network}.

We build on these results to define our orbit adjacency based node embeddings,
aiming to learn an embedding space so that nodes that tend to frequently
co-occur on a given orbit adjacency are close in the embedding space.
We define the graphlet-orbit PMI matrix, GOPMI, as: 
\begin{equation}
	GOPMI = log \left( vol(A_{o_{i \scalebox{.8}{-} j}})  D^{-1}_{row} A_{o_{i \scalebox{.8}{-} j}} D^{-1}_{col}) \right) - log(b),
\end{equation}
where \(A_{o_{i \scalebox{.8}{-} j}}\) is an orbit adjacency matrix,
\(D_{row}\) is the diagonal matrix with the row sums of \(A_{o_{i
\scalebox{.8}{-} j}}\) and \(D_{col}\) is the diagonal matrix with the column
sums of \(A_{o_{i \scalebox{.8}{-} j}}\).

As we have shown that the power of the adjacency matrix implicitly
captures a linear combination of the orbit adjacency matrices, to serve as a
baseline in our further experiments, we also define the random walk PMI matrix, RWPMI, as:
\begin{equation}
	RWPMI = log \left( vol(A^p)  D^{-1}_p A^p D^{-1}_p  \right) - log(b),
\end{equation}
where \(p\) is a power of the adjacency matrix of the network and \(D_p\) is the diagonal matrix with the sum of rows from \(A^p\).

Finally, following the work of Qiu \emph{et al.}, we apply SVD to factorise the
non-negative entries of the GOPMI and RWPMI matrices to compute our orbit
adjacency based and random walk based node embeddings. Like Qiu \emph{et al.}, we
set parameter \(b\) to 1.

\section{Data}
\label{sec:data}

To show the relevance of our methods for real-world networks across a wide
range of domains, we selected the following six networks (see Appendix Section \ref{sub:net_stats} for an overview of network statistics):
\begin{itemize} 
   \item Airports-USA: We
	use the Airports-USA network from \cite{ribeiro2017struc2vec}, in which nodes represent airports, edges
	represent flights between airports and node-labels represent airport activities.
  \item Amazon Computers: We
	use the Amazon Computers network from \cite{mcauley2015image, shchur2018pitfalls}, in which nodes represent goods, edges
	represent frequently bought together goods and node-labels represent
	product categories.  
   \item Flickr: We use the Flickr network from \cite{huang2017label, yang2020scaling}, in which nodes represent users,
	edges represent following relationships and node-labels represent user
	interests.  
   \item PubMed: We use the PubMed network from \cite{sen2008collective}, in which nodes represent papers, edges
	represent citations between papers and node-labels represent fields of study.
    \item Co-authorship network:
	We use the Coauthor-CS network from \cite{shchur2018pitfalls}, in which nodes represent authors, edges
	represent co-authorships and node-labels represent fields of study.
  \item Protein-Protein Interactions (PPI): We use the experimentally validated human PPI network from
	BioGRID (version 4.4.228) \cite{oughtred2021biogrid}, in which nodes represent proteins and edges represent
	bindings between the corresponding proteins. We filter for interactions captured by
	Yeast-2-Hybrid and Affinity Capture-MS methods. As node class-labels, we use
	biological process (BP) annotations from the Gene Ontology (GO)
	database (collected on 2024-02-09) \cite{ashburner2000gene, gene2023gene}. The BP annotations are organised
	as a directed acyclic rooted tree in which nodes represent BP terms and
	edges represent ``is a'' relationships. 
	We use the relatively generic 13 BP terms one hop
	away from the root node as node class-labels.  

\end{itemize}
Apart from the PPI network, al data has been collected via the PyTorch geometric data API \cite{paszke2019pytorch}.

\section{Empirical evaluation of the value of the topological information unseen by random walks}
\label{sec:evaluation}

Having shown that random walks capture a mixture of topological information in
terms of orbit adjacencies, but that they also miss many orbit adjacencies, we
empirically evaluate the value of the orbit adjacencies unseen by
random walks and the value of disentangling the orbit adjacencies.

We consider a multi-class node label prediction task, by applying our orbit
adjacency-based node embeddings (see Section \ref{sec:embeddings}) for six real-world
networks of various sizes and domains (see Section \ref{sec:data}). For each
network, we compute all 28 orbit adjacency based node embeddings, using the
number of different class labels as the dimension of the embedding space. Then,
for each network, class label and orbit adjacency, we perform a 10-fold
stratified cross-validation in which we train a one-vs-rest linear support
vector machine (SVM), each time randomly sampling 80\% of the nodes as our
training set and the remaining 20\% as our test set. We sample in such a way so
that the relative class frequencies are approximately preserved in each train
and test set. We measure the performance over the 10 runs in terms of micro and
macro averaged F1 score.  As the results are highly similar, we focus on the
results in terms of the average micro F1 score in the manuscript and those
based on the average macro F1 score in the Appendix. 
As topology-function relationship has been demonstrated \cite{milenkovic2008uncovering,Milenkovic_2010,Yaveroglu_2014_revealing}, 
we retain the result of the best performing adjacency for each class label, 
showing the topology characteristic to a specific class label.
We compare
our results to embeddings based on the three random walk adjacencies \(A^1, A^2
\text{ and } A^3,\) as we have shown that these matrices are linear combinations
of subsets of our orbit adjacencies.  Also, we compare to DeepWalk embeddings for
walks up to length three, as it is a widely used random walk based embedding
method and can be seen as a method that combines the overall contribution of
all walks up to a given length. 

\begin{figure}[h]
	\centering
	\includegraphics[width=.73\linewidth]{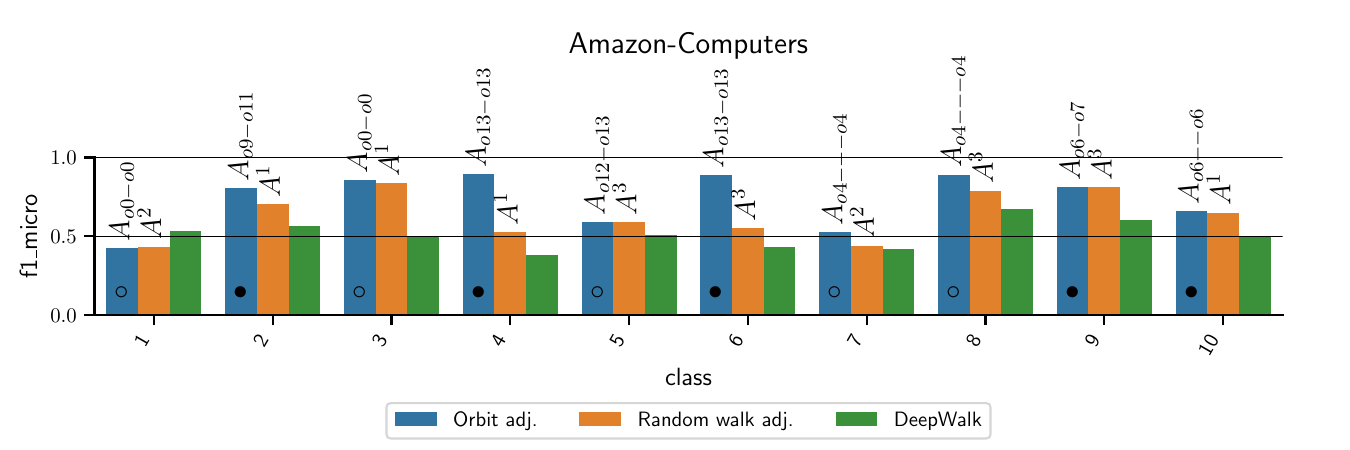}
	\caption{{ \bf Node-label prediction accuracy for each class label in the Amazon-Computer network.}  We show for each class label (x-axis) and each model (legend), the node-label prediction accuracy achieved by the best performing underlying adjacency (detailed above each bar), measured using the micro averaged F1 score. In the case of orbit adjacency embeddings, $\circ$ and \(\bullet\) on the bar indicate if the corresponding best performing orbit adjacency is seen or unseen by random walks up to length three, respectively.}
	\label{fig:best_adjacency}
\end{figure}

We detail our results for the Amazon-Computer network in Fig.
\ref{fig:best_adjacency}. For this network, we observe that the best orbit
adjacency outperforms or ties the best random walk adjacency and DeepWalk for
nine out of ten classes when considering the micro averaged F1 score.
Importantly, for five of the nine classes where orbit adjacencies outperform or
tie random walk adjacency and DeepWalk, the best performing orbit adjacencies
are not seen by random walks, illustrating the value of the topological
neighbourhood information that is captured by orbit adjacencies, but unseen by
random walks. There are also four classes for which the orbit adjacencies that
outperform or tie random walk adjacency and DeepWalk are seen by random walks,
which illustrates the value of disentangling the different types of topological
neighbourhoods and being able to select the most relevant ones for a given
task.

\begin{figure}[h]
	\centering
		\begin{tabular}{cc}
			\includegraphics[width=0.45\linewidth]{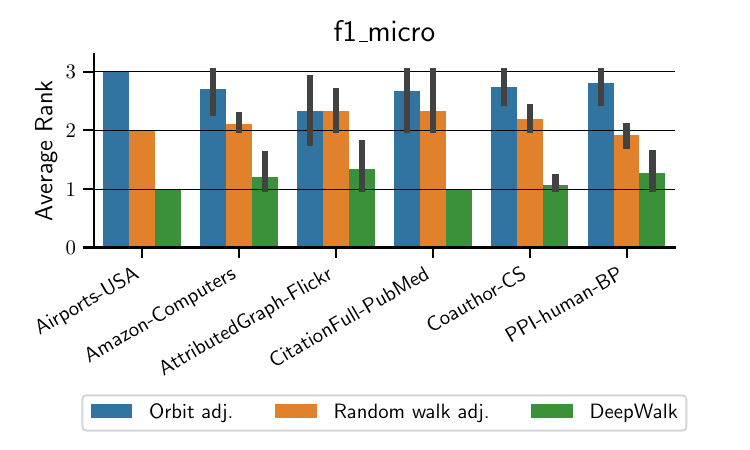} &
			\includegraphics[width=0.45\linewidth]{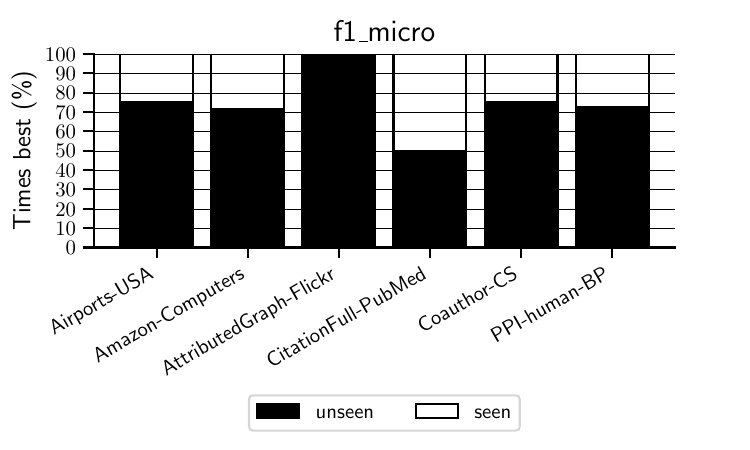} \\
		\end{tabular}
		\caption{{\bf Performance evaluation of the best type of adjacency across six different networks.}
			On the left, for each of the six networks, we show the average rank of the best performing adjacency for each embedding strategy, measured by using the micro averaged F1 score. 
		The error bars represent the 95\% confidence interval of the average rank, computed by using bootstrapping. On the right, we show the proportion of times the best performing orbit adjacency is unseen by random walks up to length three, when the orbit adjacency outperforms random walk adjacency and DeepWalk, measured by using the micro averaged F1 score.}
	\label{fig:summary}
\end{figure}

To show that our results hold for a wide range of networks, we summarize the
overall performance of a given embedding strategy (either based on orbit
adjacencies, random walk adjacencies or DeepWalk using random walks of lengths
up to 3) as follows. For each network, class label and embedding strategy, we
only consider the best-performing adjacency matrix according to the micro
averaged F1 score and rank each of the three embedding strategies based on the
micro averaged F1 score of their best-performing adjacency matrices. Then, we measure the
overall performance of a network embedding strategy with the average rank
over all class labels of its best-performing adjacency matrix, which we present
for all networks in Fig. \ref{fig:summary}. For all six networks, the average
rank of orbit adjacency is higher than or equal to that of random walk adjacency and
DeepWalk. At the same time, we observe that for
the classes where orbit adjacency outperforms random walk adjacency and
DeepWalk, 74\% of the times the best performing orbit adjacency is
unseen by random walks up to length three, on average, across all networks.
Again, this illustrates the value of the topological neighbourhood information
that is captured by orbit adjacencies, but unseen by random walks.

In the Appendix, we demonstrate that our results are robust with respect to the
size of the training set by varying the size of the training set from 10\% to
90\%, using either the micro or macro averaged F1 score (see Appendix Section
\ref{sub:training_size_robustness}).  Also, we demonstrate that our results are
robust with respect to the choice of the embedding space dimension by varying
the dimension from one quarter of the number of class labels to four times the
number of class labels, using either the micro or macro averaged F1 score (see
Appendix Section \ref{sub:dimension_robustness}).

\section{Limitations}
\label{sec:limitations}

First, in this work, we empirically show the importance of the topological
information captured by our orbit adjacencies in the context of network
embedding by considering a multi-class label prediction task. However, network
embeddings are used for many different tasks, e.g., link prediction and
community detection, for which future research would have to show the value of
our orbit adjacencies.

Second, in our multi-class label prediction task, we train a separate SVM for
each class label and orbit adjacency to show that there is value in considering
the topological neighbourhood information currently ignored by random-walk
embedding approaches and that there is value in disentangling the different
topological neighbourhood information which is implicitly mixed by current
random walk based embedding methods. We do not claim this approach to be the best and
most efficient way to build a predictor. Rather, our results open up the
research questions of how to best jointly mine these topologies, or how to
select them \emph{a priori} for a given task.

Third, to enable orbit adjacency-based analysis for large real networks, we
release our GRADCO software. However, the space and time complexities of GRADCO
are both in \(\bigO{n^3}\), where \(n\) is the number of nodes in the network.
So despite our efficient implementation, detailed in Appendix Section
\ref{sub:gradco}, GRADCO may not be suitable for huge networks with millions of
nodes, unlike the state-of-the art random walk based bedding methods. As the
current scalability of GRADCO does not take away from our results, we believe
that this invites new research to include orbit adjacency information in more
scalable sampling-based methods.

Fourth, in this work, we considered the topological neighbourhood information
captured by random walks of length up to three (covering subgraphs having up to
four nodes). Of course, random walk based embeddings can consider longer random
walks and thus, implicitly, consider topological neighbourhood information
based on larger subgraphs.  However, as we highlighted in section
\ref{randomwalkneighborhood}, these longer random walks having length \(l\geq
3\) will still miss many of the topological information captured by the up to
\((l+1)\)-node subgraphs. Furthermore, we note that real networks are usually
small world, i.e., the mean shortest path length between nodes is proportional
to the logarithm of the number of nodes in the network
\citep{Watts_1998_small-world}, so in practice one does not need to consider
large subgraphs to successfully mine these data.

Finally, in this work, we define orbit adjacency for unweighted, undirected
networks. Graphlet-based methods have been generalised for many different types
of networks, e.g., directed networks \cite{sarajlic2016graphlet}, 
weighted networks \cite{Sergio-probabilistic-2020}, temporal
networks \cite{paranjape2017motifs} and hyper networks
\cite{gaudelet2018higher}. As such, orbit adjacency can also be generalised to
these and other types of networks, allowing for the development of novel
network embedding and analysis methods that better capture and exploit the
topological neighborhoods of the nodes than the methods based on random walks.

\section{Conclusion}%
\label{sec:Conclusion}

Although random walks are widely used for mining networks, the topological
neighbourhood information that they implicitly capture is poorly understood. In
this work, we introduce orbit adjacency, which quantifies the adjacencies of two
nodes as their frequency of co-occurrence on a given pair of graphlet orbits.
We use orbit adjacency to explicitly characterize the topological neighbourhood
information captured by random walks, by mathematically proving that random walks
on up to \(k\) nodes capture only a sub-set of all possible orbit
adjacencies for up to \(k\)-node graphlets. To enable orbit adjacency-based
analysis of real networks, we develop GRADCO, which exhaustively computes all
28 orbit adjacency matrices for up to four-node graphlets; for large networks of
around 20,000 nodes, it computes them in minutes. To show the practical value of the orbit
adjacencies unseen or entangled by random walks, we develop our PMI based orbit
adjacency embeddings and apply them in a multi-class node label prediction task
for six real-world networks, comparing to random walk based embeddings. We find
that orbit adjacencies, which include those unseen by random walks, outperform
random walk-based adjacencies, demonstrating the importance of the inclusion of
the topological neighborhood information that is unseen by random walks. Our
results open up multiple research questions, e.g.: (a) the inclusion of orbit
adjacency information in more scalable, sampling-based architectures; (b) how
to best jointly mine these different types of topological neighborhoods; and
(c) how to select the specific topological neighborhood \emph{a priori} that works
best for a given task.

{
\small

\bibliographystyle{unsrt}
\bibliography{biblio}

}


\newpage

\newpage
\appendix

\section{Appendix / supplemental material}

\subsection{Network statistics}
\label{sub:net_stats}

Table \ref{tab:net_stats} shows the statistics of the networks used in our experiments.
Things worth nothing are:
\begin{itemize}
	\item Our GRADCO is able to compute the orbit adjacencies for any of these networks in less than five minutes.
	\item The $90^th$ percentile of shortest paths is between 3 and 8 for these networks, supporting our claim that to capture the local neighbourhood structure in real networks, considering up to 4-node graphlets is sufficient.
	\item For the PPI-human-BP network, only 50\% of the nodes have at least one class label. This is important as we only consider nodes with at least one class label in our experiments in Section \ref{sec:evaluation}.
	
\end{itemize}

\begin{table}[h]
	\centering
\caption{{\bf Network statistics.} The table shows the number of nodes, edges, density, the $90th$ percentile of shortest paths, the number of classes, the percentage of nodes that have at least one class label (i.e., node coverage) and the runtime of GRADCO for each of our networks.}
\begin{tabular}{lrrrrrrrr}
\toprule
Network & Nodes & Edges & \makecell{Density \\ (\%)} & \makecell{$90th$ \\ perct. \\ Shortest \\ paths} & Classes & \makecell{Node \\coverage \\ (\%) }& \makecell{GRADCO \\runtime \\ (sec)} \\
\midrule
PPI-human-BP & 19,430 & 629,784 & 0.334  & 3 & 13 & 50 & 261 \\
Coauthor-CS & 18,333 & 81,894 & 0.05  & 7 & 15 & 100 & 130 \\
Amazon-Computers & 13,471 & 245,861 & 0.27  & 4 & 10 & 100 & 72 \\
AttributedGraph-Flickr & 7,575 & 239,738 & 0.84  & 3 & 9 & 100 & 50 \\
Airports-USA & 1,190 & 13,599 & 1.92  & 4 & 4 & 100 & 0.54 \\
CitationFull-PubMed & 19,717 & 44,324 & 0.023  & 8 & 3 & 100 & 157 \\
\bottomrule
\end{tabular}
\label{tab:net_stats}
\end{table}


\FloatBarrier

\subsection{Unseen orbit adjacencies}
\begin{sidewaysfigure}[htpb]
	\centering
	\begin{subfigure}[]{\linewidth}
	\begin{center}
	\includegraphics[width=.25\linewidth]{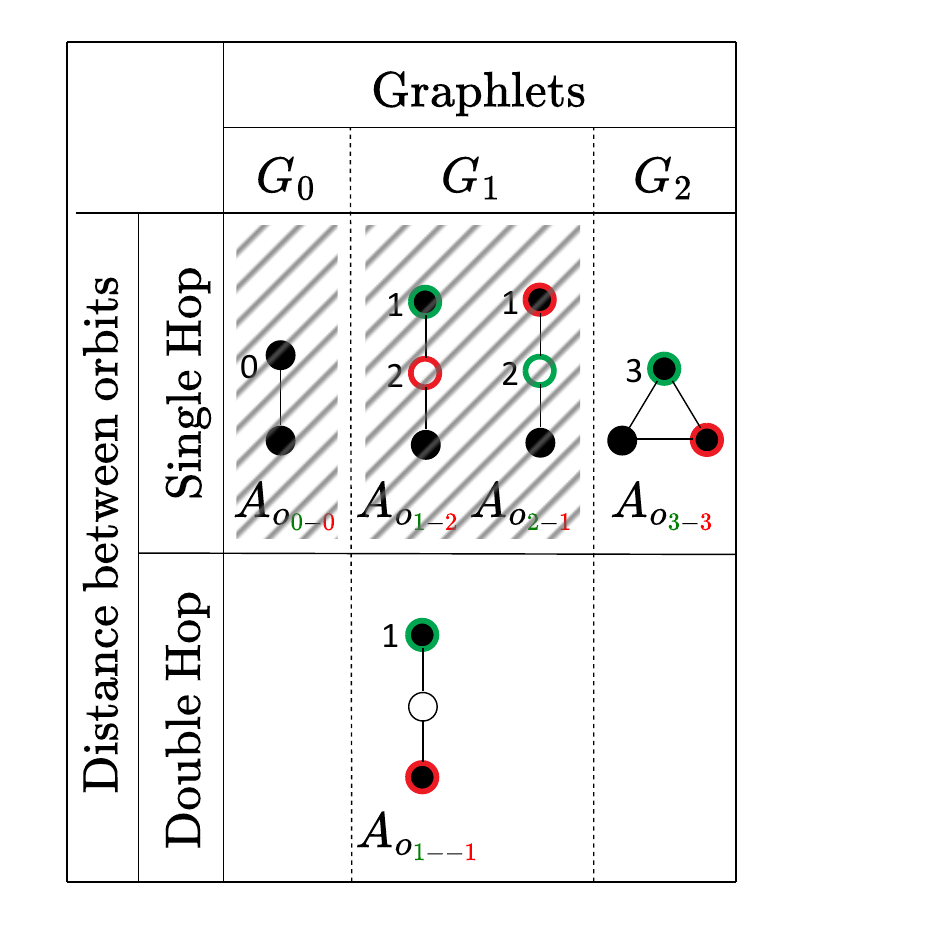}\\ 
	\end{center}
	\caption{ The orbit adjacencies seen and unseen by random walks of length two.}
	\hfil
	\end{subfigure}
	\hfil
	\begin{subfigure}[]{\linewidth}
	\begin{center}
	\includegraphics[width=.95\linewidth]{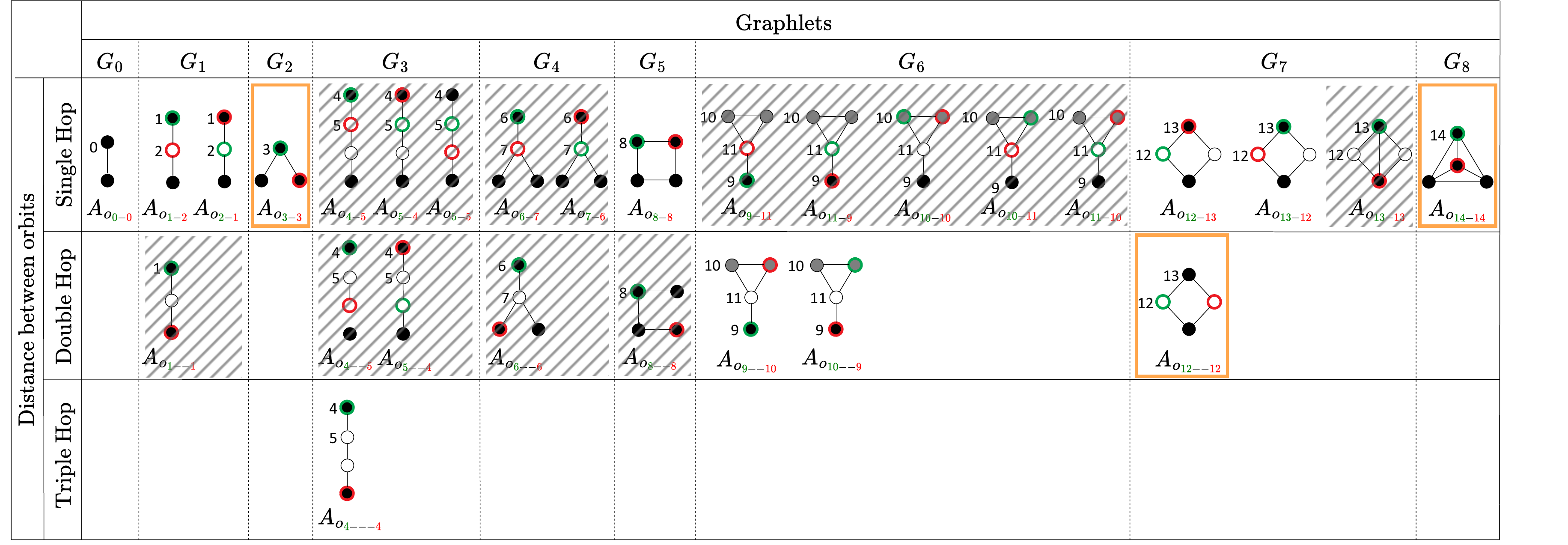}
	\end{center}
	\caption{ The orbit adjacencies seen and unseen by random walks of length three.}
	\end{subfigure}
	\caption{The orbit adjacencies seen and unseen by random walks of length two and three. The orbit adjacencies that are seen unseen by random walks of a given length are hatched. The orbit adjacencies that are seen multiple times by random walks of a given length are lined with a solid orange line.}
	\label{fig:blind_adjacencies}
\end{sidewaysfigure}

In Fig. \ref{fig:blind_adjacencies}, we illustrate all the orbit adjacencies that are seen and unseen by random walks of length two and three.

\newpage

\subsection{Detailed performance results}

\begin{figure}[h]
	\centering
	\begin{tabular}{c}
		\includegraphics[width=0.72\linewidth]{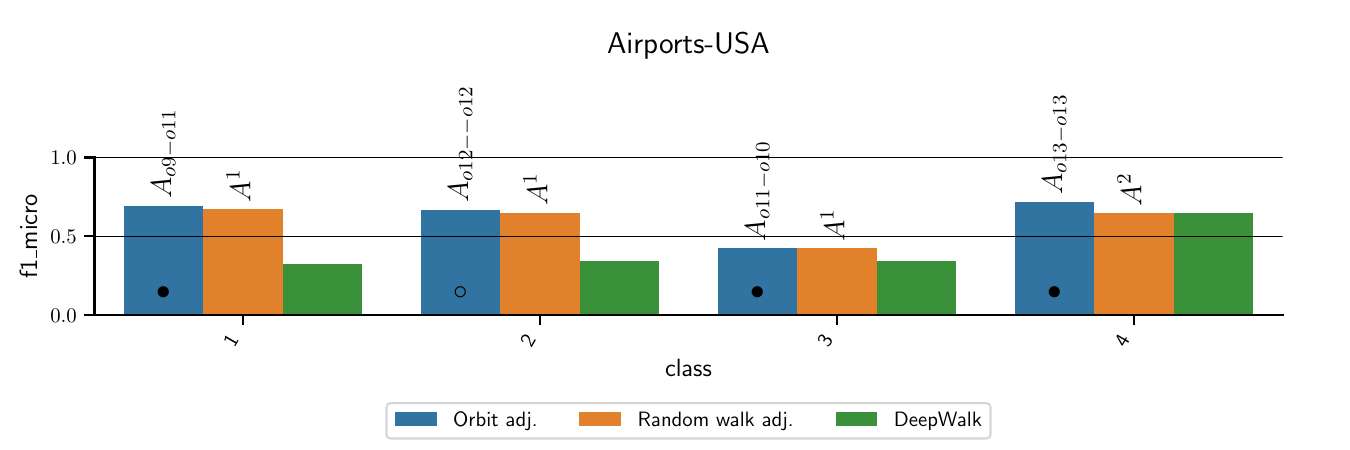} \\
		\includegraphics[width=0.72\linewidth]{figs/best_adjacency_per_class_Amazon-Computers_f1_micro_80.pdf} \\
		\includegraphics[width=0.72\linewidth]{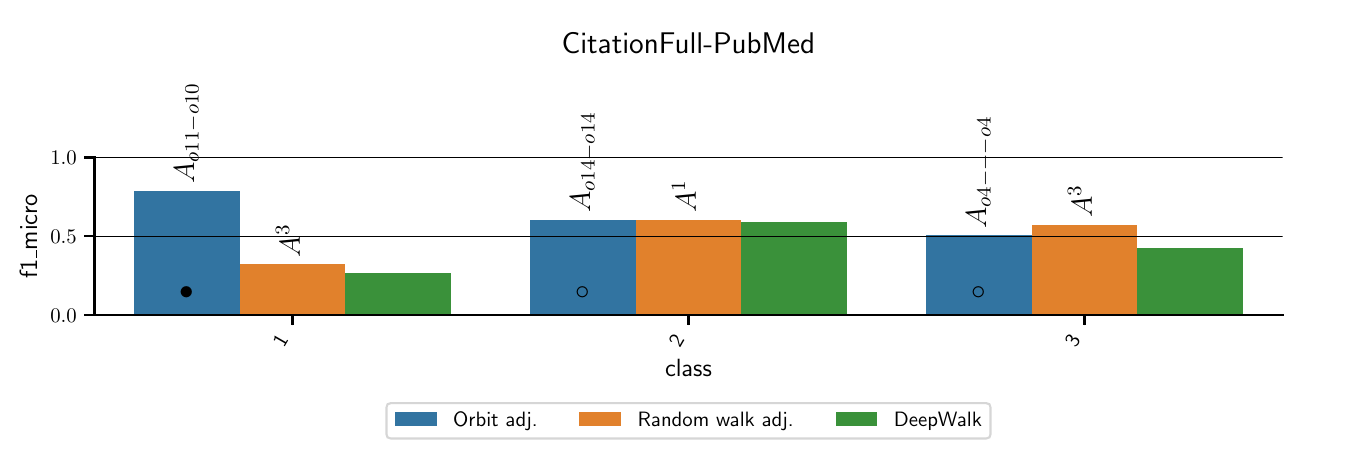} \\
		\includegraphics[width=0.72\linewidth]{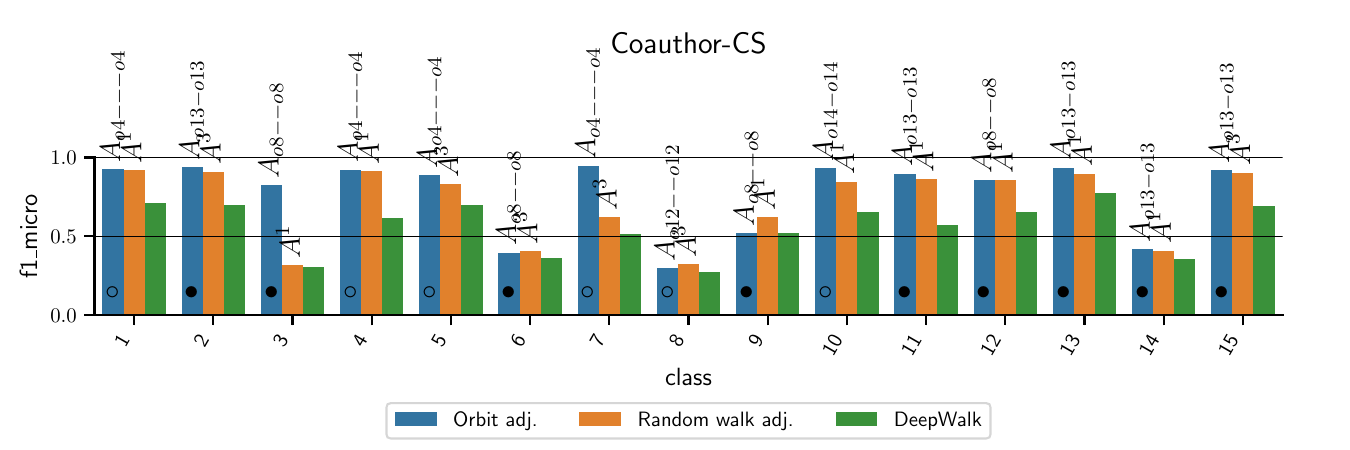} \\
		\includegraphics[width=0.72\linewidth]{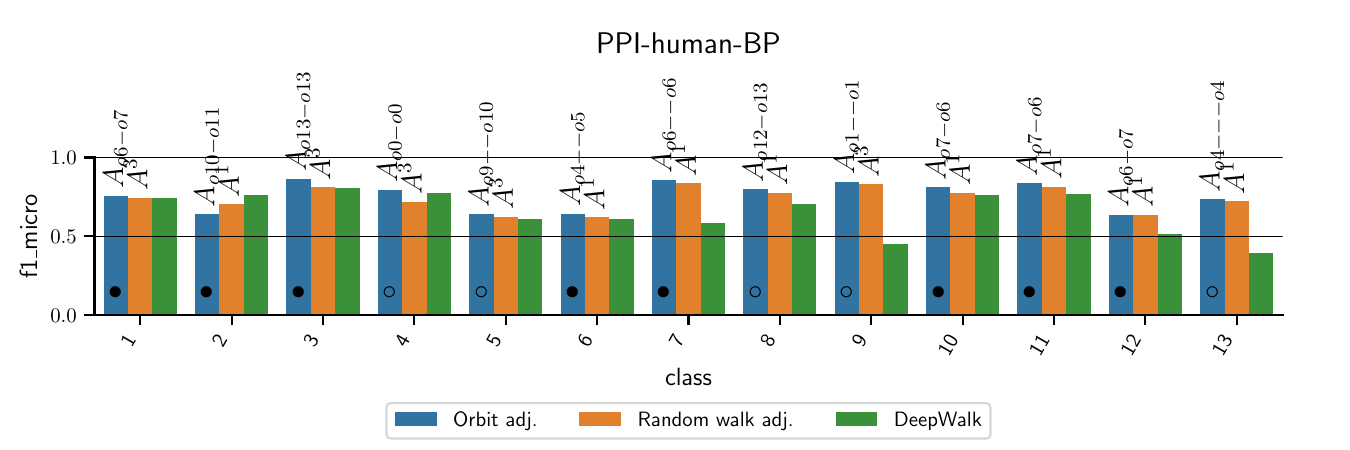} \\
		\includegraphics[width=0.72\linewidth]{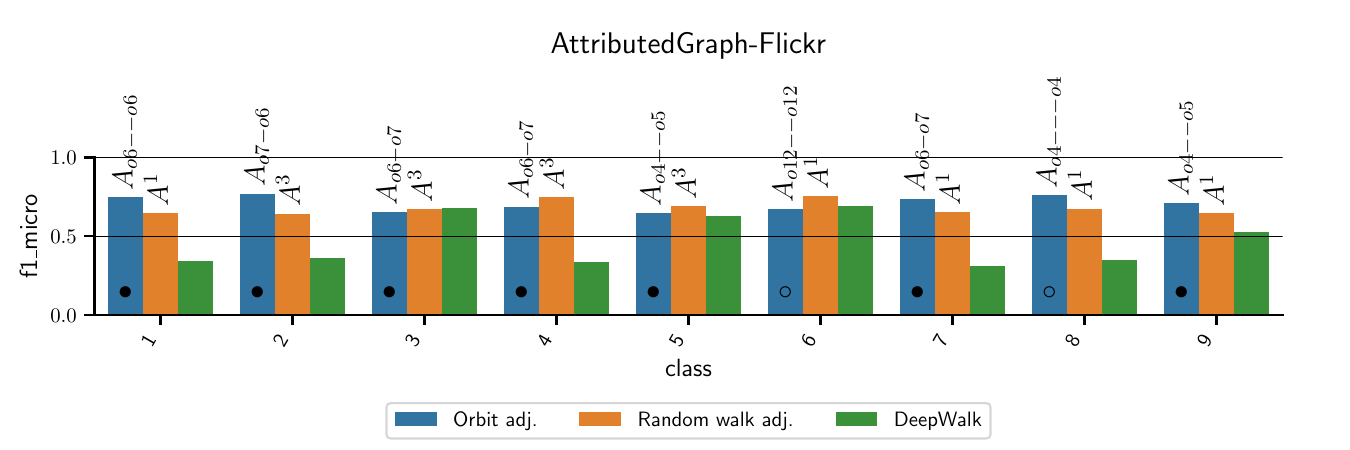} \\
	\end{tabular}
	\caption{{ \bf Node-label prediction accuracy for each class label for our six networks using the micro averaged F1 score.} 
	For each of our six networks (top to bottom), we show for each class label (x-axis) and each model (legend), the node-label prediction accuracy achieved by the best performing underlying adjacency (detailed above each bar), measured using the micro averaged F1 score. In the case of orbit adjacency embeddings, a $\circ$ and \(\bullet\) on the bar indicate if the corresponding best performing orbit adjacency is seen or unseen by random walks up to length three, respectively.}
	\label{fig:detailed_performance_micro}
\end{figure}

\begin{figure}[h]
	\centering
	\begin{tabular}{c}
		\includegraphics[width=0.72\linewidth]{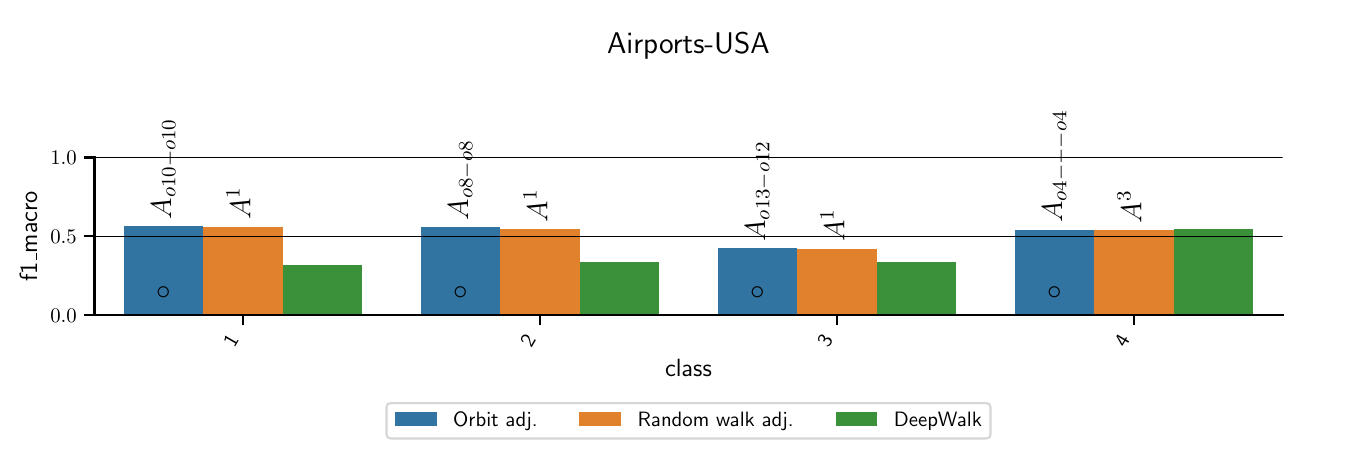} \\
		\includegraphics[width=0.72\linewidth]{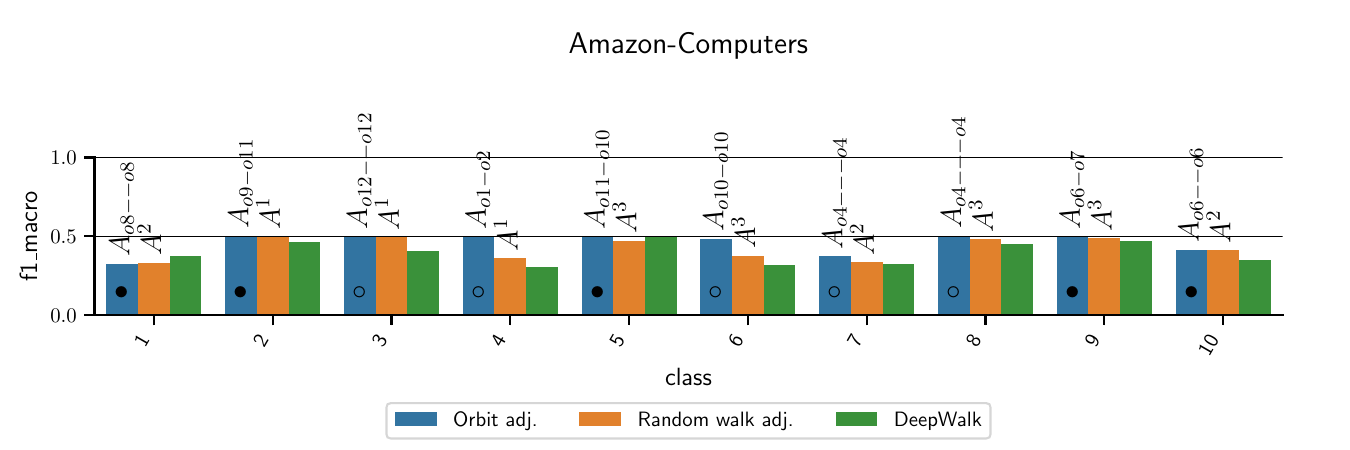} \\
		\includegraphics[width=0.72\linewidth]{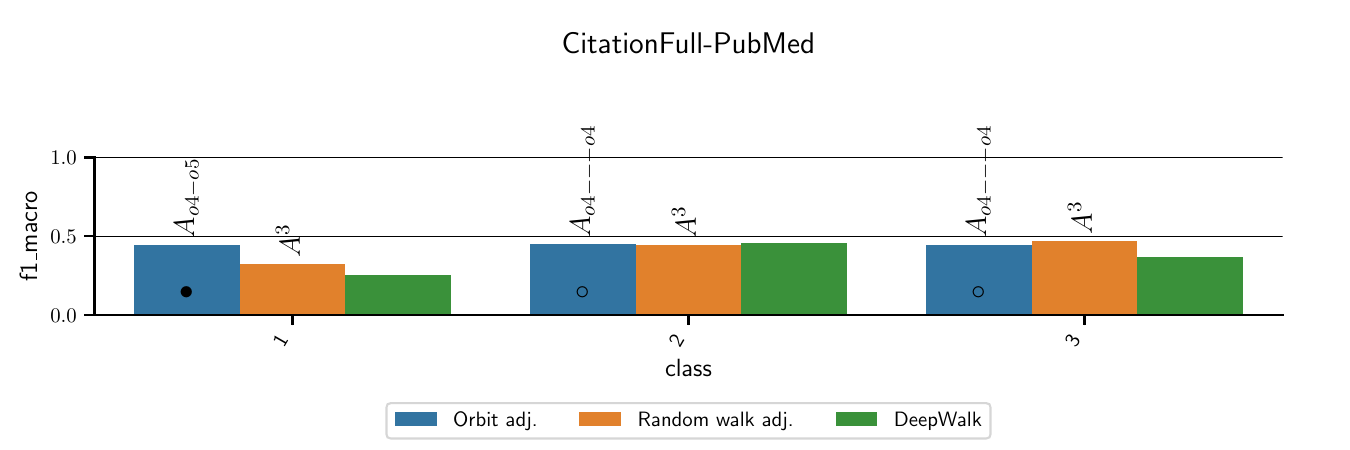} \\
		\includegraphics[width=0.72\linewidth]{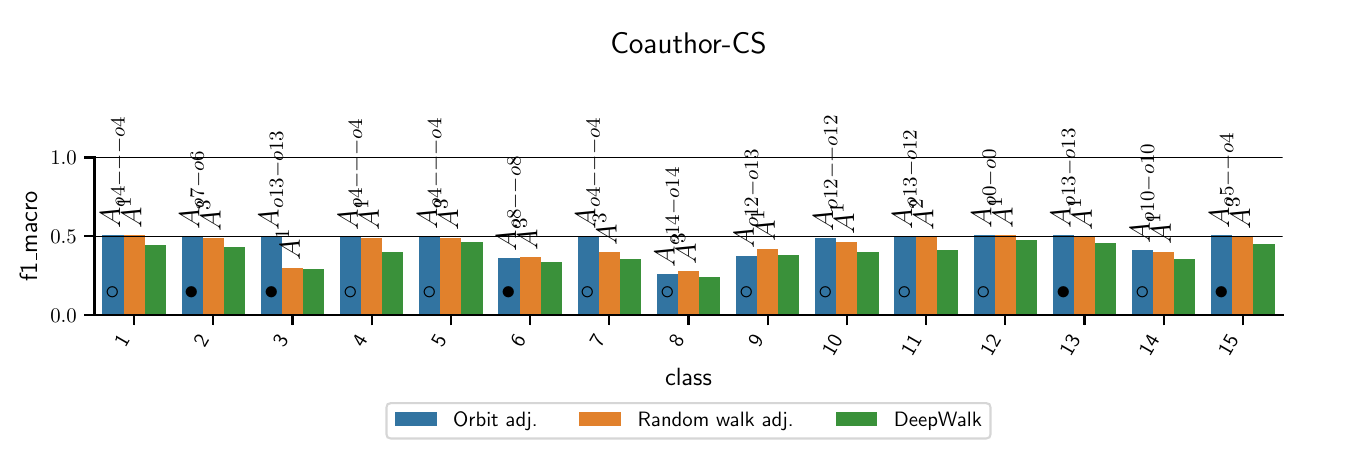} \\
		\includegraphics[width=0.72\linewidth]{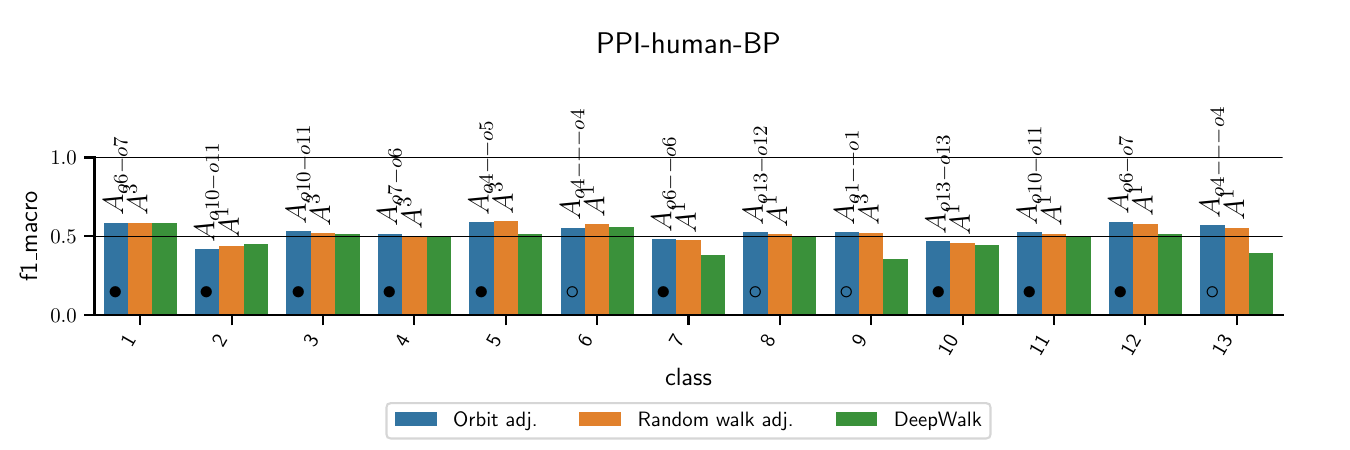} \\
		\includegraphics[width=0.72\linewidth]{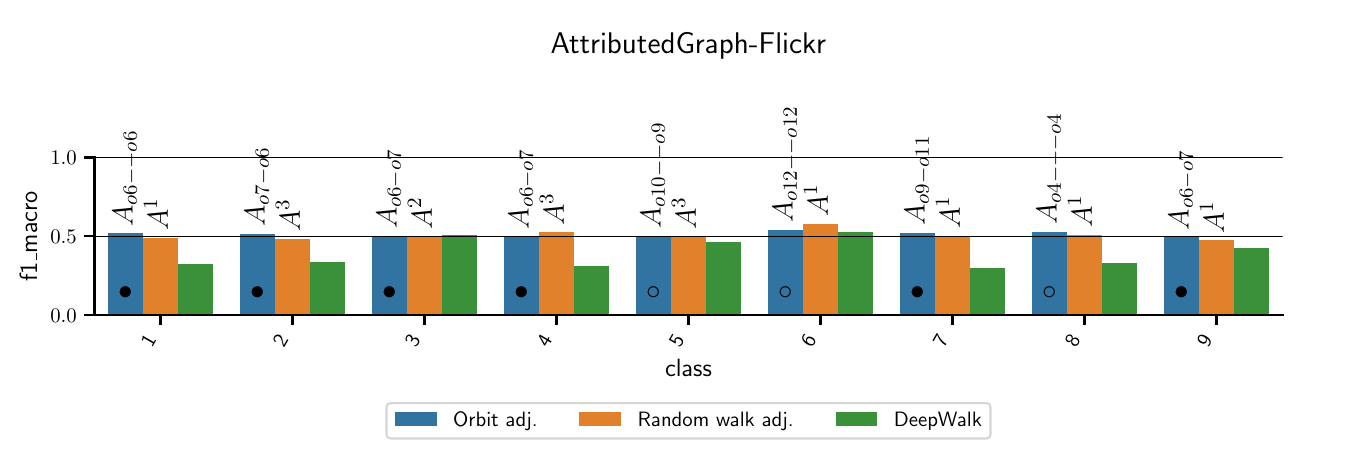} \\
	\end{tabular}
	\caption{{ \bf Node-label prediction accuracy for each class label for our six networks using the macro averaged F1 score.} 
	For each of our six networks (top to bottom), we show for each class label (x-axis) and each model (legend), the node-label prediction accuracy achieved by the best performing underlying adjacency (detailed above each bar), measured using the macro averaged F1 score. In the case of orbit adjacency embeddings, a $\circ$ and \(\bullet\) on the bar indicate if the corresponding best performing orbit adjacency is seen or unseen by random walks up to length three, respectively.}
	\label{fig:detailed_performance_macro}
\end{figure}

In Section \ref{sec:evaluation}, to empirically show the importance of
the orbit adjacencies that are unseen by random walks and the value of disentangling them,
we considered a multi-class node label prediction task, applying our orbit
adjacency-based node embeddings for six real-world networks of various sizes
and domains. In the main paper, we showed our results for the Amazon-Computer
network using the averaged micro F1 score. Here, we show the detailed
class-prediction results for all six networks using both the micro and macro
averaged F1 scores in Figures  \ref{fig:detailed_performance_micro} and
\ref{fig:detailed_performance_macro}, respectively.

\FloatBarrier

\newpage
\subsection{Robustness with respect to the size of the training set}%
\label{sub:training_size_robustness}

In Section \ref{sec:evaluation}, to empirically show the importance of the
orbit adjacencies unseen by random walks and the value of disentangling them,
we considered a multi-class node label prediction task, applying our orbit
adjacency-based node embeddings for six real-world networks of various sizes
and domains. We showed that in general across all six networks (1) the best
orbit adjacency per class outperformed or tied the best random walk adjacency and
DeepWalk using the micro averaged F1 score and (2) that for the classes where
orbit adjacency outperforms random walk adjacency and DeepWalk, 74\% of the
times the best performing orbit adjacency is unseen by random walks up to
length three. In our experiment, we only considered the micro averaged F1 score
and fixed the size of the training set to 80\% of the data. Here, to show that
both our results are robust with respect to the size of the training set and the
performance measure used, we repeat the same experiment, varying the size of
the training set from 10\% to 90\%, whilst considering both the micro averaged
and macro averaged F1 score.

\begin{figure}[h]
	\centering
		\begin{tabular}{cc}
			\includegraphics[width=0.5\linewidth]{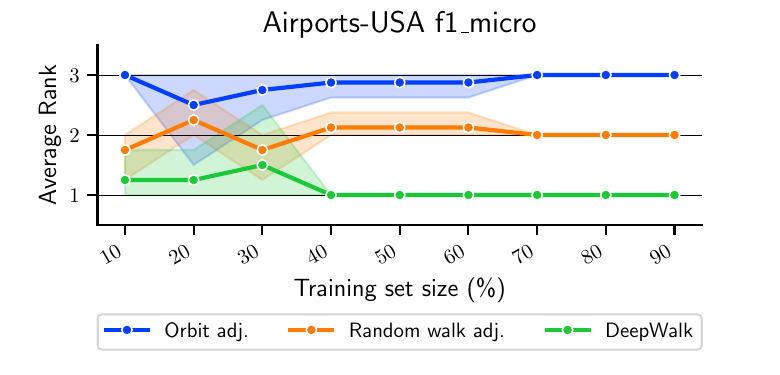} &
			\includegraphics[width=0.5\linewidth]{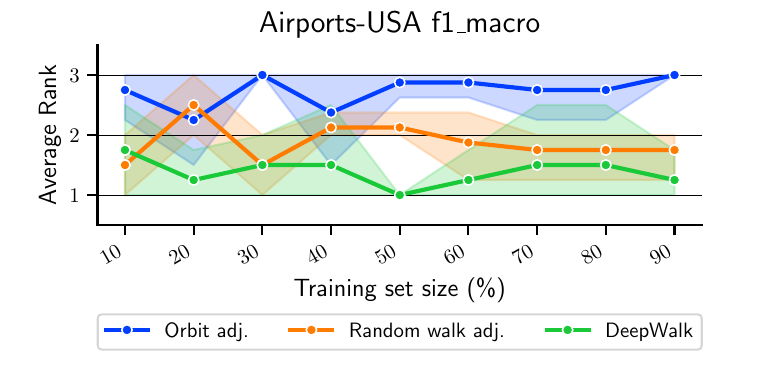} \\
			\includegraphics[width=0.5\linewidth]{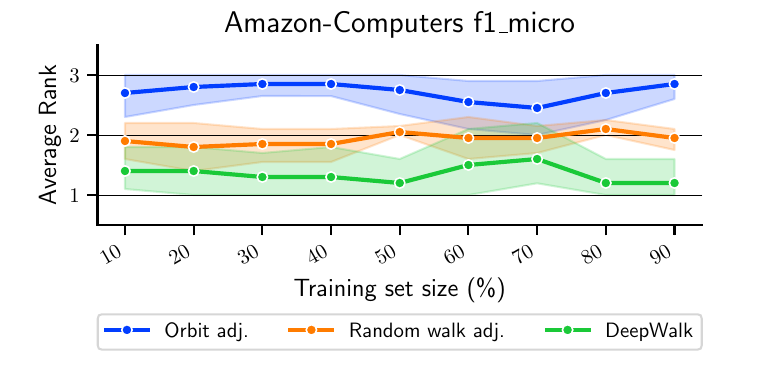} &
			\includegraphics[width=0.5\linewidth]{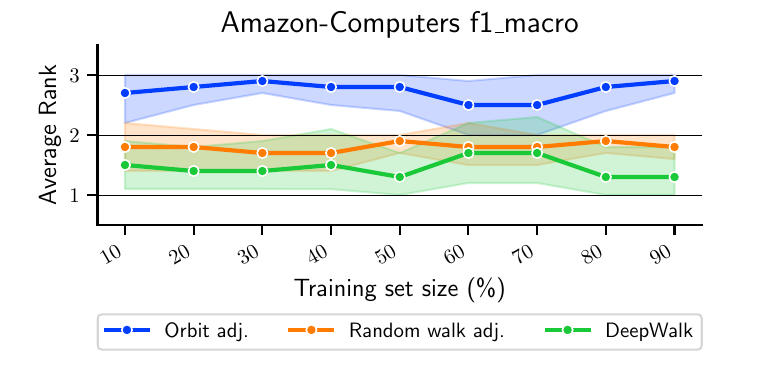} \\
			\includegraphics[width=0.5\linewidth]{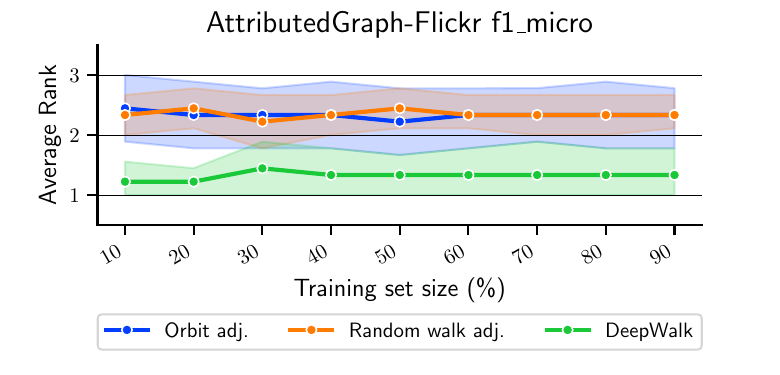} &
			\includegraphics[width=0.5\linewidth]{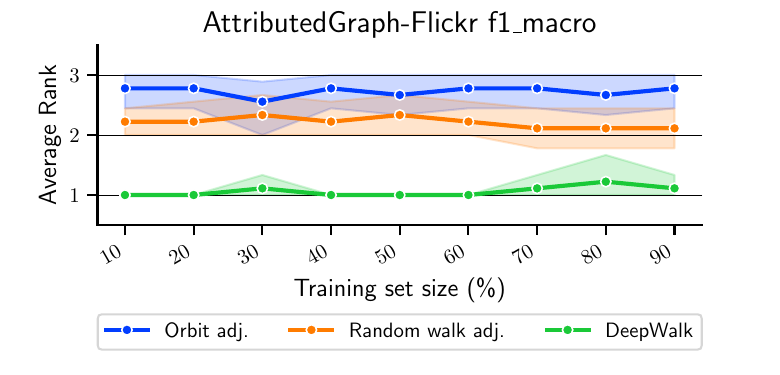} \\
			\includegraphics[width=0.5\linewidth]{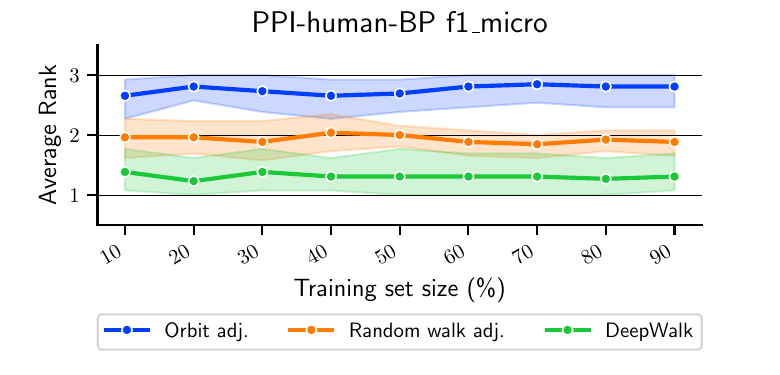} &
			\includegraphics[width=0.5\linewidth]{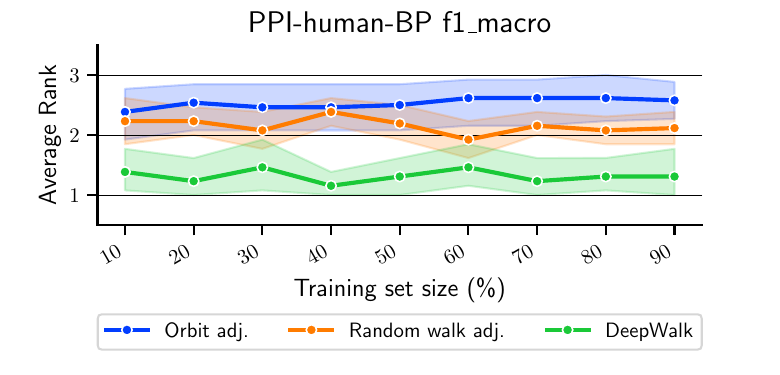} \\
			\includegraphics[width=0.5\linewidth]{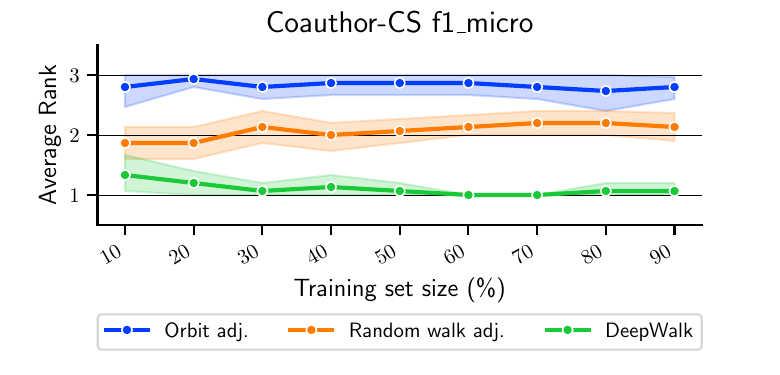} &
			\includegraphics[width=0.5\linewidth]{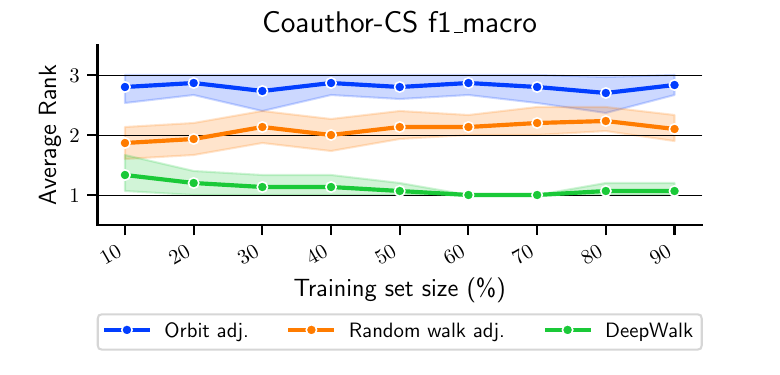} \\
		\end{tabular}
	\caption{{\bf Impact of the training set size on the ranking of the performance of the best type of adjacency across six different networks.}
			For each of our six networks (top to bottom), varying the size of the training set from 10\% to 90\% (x-axis), we show the average rank of the best performing adjacency for each embedding strategy, measured using the micro averaged F1 score (left) and macro averaged F1 score (right).The error bars represent the 95\% confidence interval of the average rank, computed using bootstrapping.}
	\label{fig:training_size_sensitivity}
\end{figure}

To assess the robustness of result (1), we show the average rank of the best
performing adjacency across all class labels for each of our six networks in
Fig.  \ref{fig:training_size_sensitivity}.  We observe that across all
networks, the average rank of the best orbit adjacency is higher or equal to
that of the best random walk adjacency and DeepWalk for both the micro and
macro averaged F1 score regardless of the size of the training set. Three
points of exceptions are: (1) the Airports-USA network for the macro averaged
F1 score and considering a training set size of 20\% and the Flickr graph for
the micro averaged F1 score, where random walk adjacency slightly outperforms
orbit adjacency. This shows that result (1) is robust with respect to the size
of the training set.

\begin{figure}[h]
	\centering
	\begin{tabular}{cc}
	\includegraphics[width=0.5\linewidth]{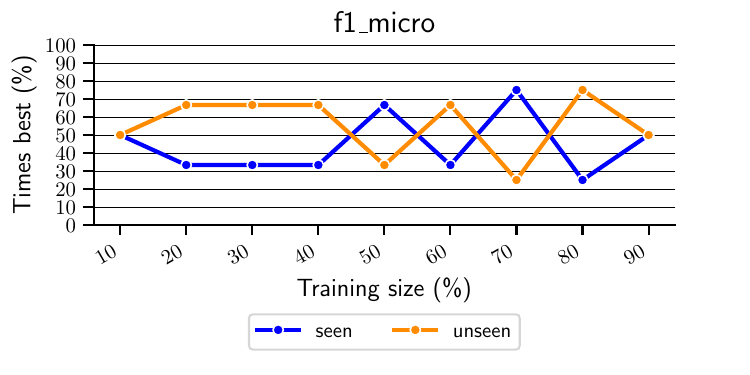} &	
	\includegraphics[width=0.5\linewidth]{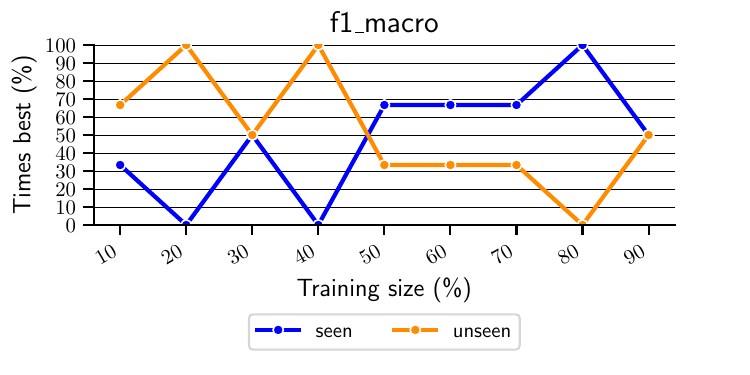} \\
	\includegraphics[width=0.5\linewidth]{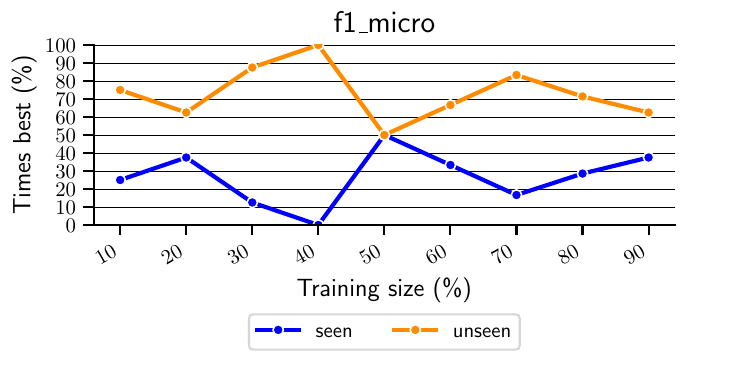} &
	\includegraphics[width=0.5\linewidth]{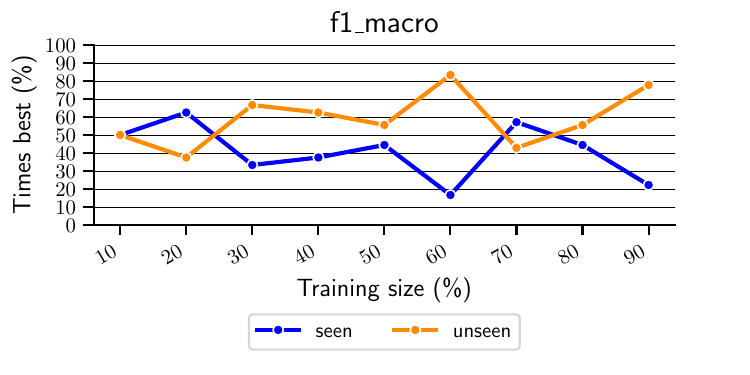} \\
	\includegraphics[width=0.5\linewidth]{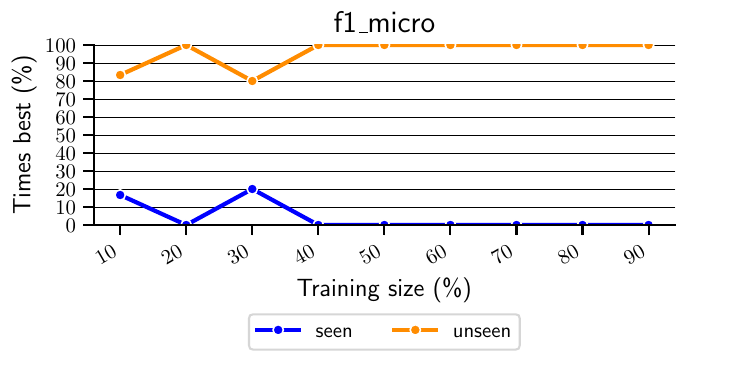} &
	\includegraphics[width=0.5\linewidth]{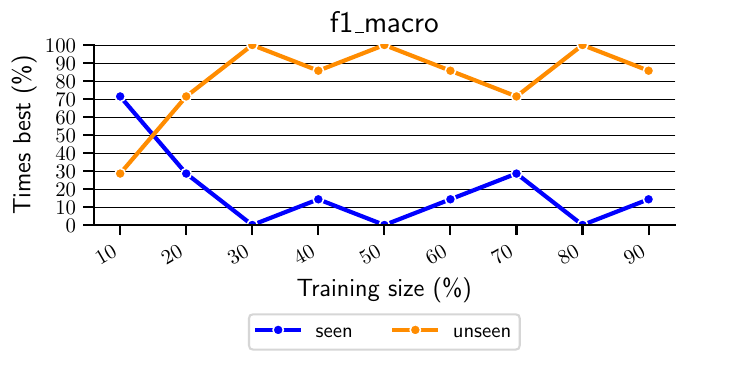} \\
	\includegraphics[width=0.5\linewidth]{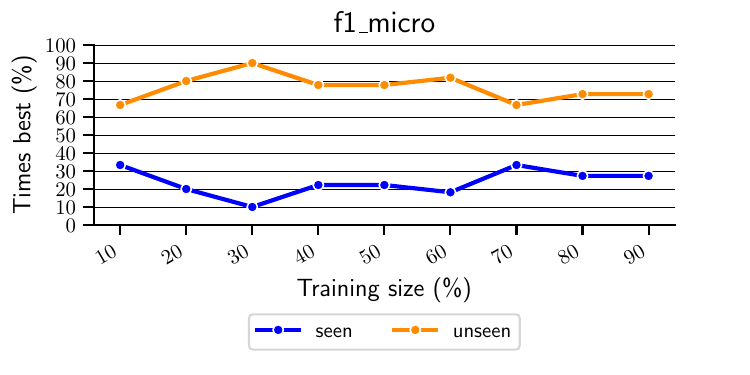} &
	\includegraphics[width=0.5\linewidth]{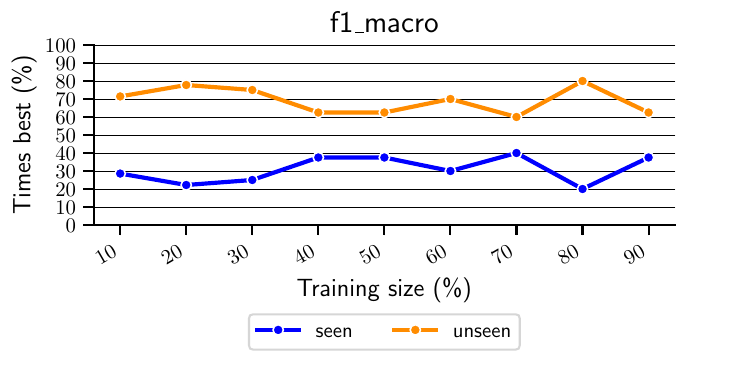} \\
	\includegraphics[width=0.5\linewidth]{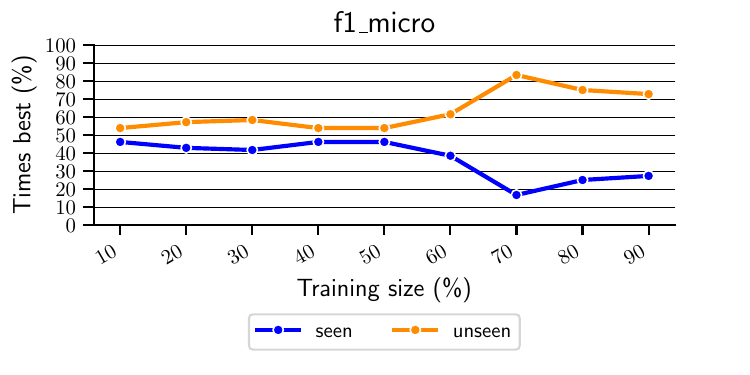} &
	\includegraphics[width=0.5\linewidth]{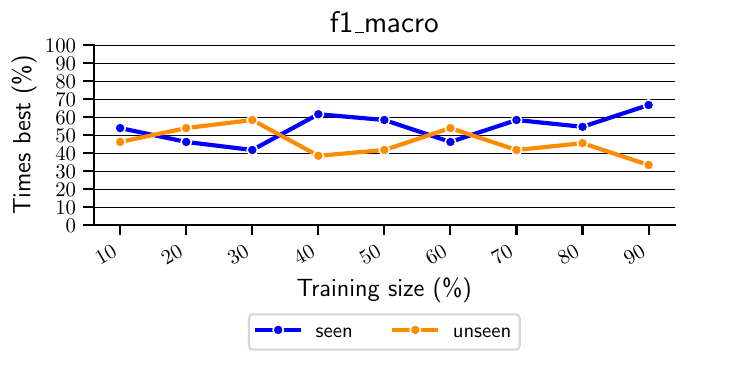} \\

	\end{tabular}
	\caption{{\bf The impact of the training set size on what proportion of the orbit adjacencies that outperform random walks adjacencies, are orbit adjacencies that are seen by random walks, across six different networks.} 
		For each of our six networks (top to bottom) we
show the proportion of times the best performing orbit adjacency is unseen by
random walks up to length three, when the orbit adjacency outperforms random
walk adjacency and DeepWalk, as measured using the micro averaged F1 score (left) and the macro averaged F1 score (right).}
	\label{fig:training_size_impact_seen}
\end{figure}

To asses the robustness of result (2), we show the proportion of times the best
performing orbit adjacency is unseen by random walks up to length three, when
the orbit adjacency outperforms random walk adjacency and DeepWalk, as measured
using the micro and macro averaged F1 score in Fig \ref{fig:training_size_impact_seen}.
On average across all networks and training set sizes, we observe that 68\%
(59\%) of the orbit adjacencies that outperform random walk adjacency and
DeepWalk are unseen by random walks up to length three, when considering the
micro (macro) averaged F1 score. This shows that result (2) is are robust with
respect to the size of the training set.

In conclusion, in general, the best orbit adjacency per class outperforms or
ties the best random walk adjacency and DeepWalk for both the micro and macro
averaged F1 score, regardless of the size of the training set. Furthermore, for
the classes where orbit adjacency outperforms random walk adjacency and
DeepWalk, the best performing orbit adjacencies are often unseen by random
walks up to length three highlighting the value of the topological
neighbourhood information that is captured by orbit adjacencies.

\FloatBarrier

\subsection{Robustness with respect to embedding dimension}%
\label{sub:dimension_robustness}


In Section \ref{sec:evaluation}, to empirically show the importance of the
orbit adjacencies unseen by random walks and the value of disentangling them,
we considered a multi-class node label prediction task, applying our orbit
adjacency-based node embeddings for six real-world networks of various sizes
and domains. We showed that in general across all six networks (1) the best
orbit adjacency per class outperformed or tied the best random walk adjacency and
DeepWalk using the micro averaged F1 score and (2) that for the classes where
orbit adjacency outperforms random walk adjacency and DeepWalk, 74\% of the
times the best performing orbit adjacency is unseen by random walks up to
length three. In our experiment, we only considered the micro averaged F1 score
and fixed the dimension of the embedding space to the number of class labels.
Here, to show that both our results are robust with respect to the dimension of
the embedding space and the performance measure used, we repeat the same
experiment, varying the dimension of the embedding space from one quarter of
the number of class labels to four times the number of class labels of any
given network, whilst considering both the micro averaged and macro averaged F1
score.

To assess the robustness of result (1), we show the average rank of the best
performing adjacency across all class labels for each of our six networks in
Fig. \ref{fig:training_size_sensitivity}.  We observe that across all
networks, the average rank of the best orbit adjacency is higher or equal to
that of the best random walk adjacency and DeepWalk for both the micro and
macro averaged F1 score regardless of the dimension of the embedding space. 
This shows that result (1) is robust with respect to the dimension of the
embedding space.

To asses the robustness of result (2), we show the proportion of times the best
performing orbit adjacency is unseen by random walks up to length three, when
the orbit adjacency outperforms random walk adjacency and DeepWalk, as measured
using the micro and macro averaged F1 score in Fig \ref{fig:training_size_impact_seen}.
On average across all networks and dimensions, we observe that 64\%
(55\%) of the orbit adjacencies that outperform random walk adjacency and
DeepWalk are unseen by random walks up to length three, when considering the
micro (macro) averaged F1 score. This shows that result (2) is are robust with
respect to the dimension of the embedding space.

In conclusion, in general, the best orbit adjacency per class outperforms or ties the best
random walk adjacency and DeepWalk for both the micro and macro averaged F1
score, regardless of the dimension of the embedding space. Furthermore, for the classes
where orbit adjacency outperforms random walk adjacency and DeepWalk, the best
performing orbit adjacencies are often unseen by random walks up to length
three, underlining the value of the topological neighbourhood information that
is captured by orbit adjacencies.

\begin{figure}[h]
	\centering
		\begin{tabular}{cc}
			\includegraphics[width=0.45\linewidth]{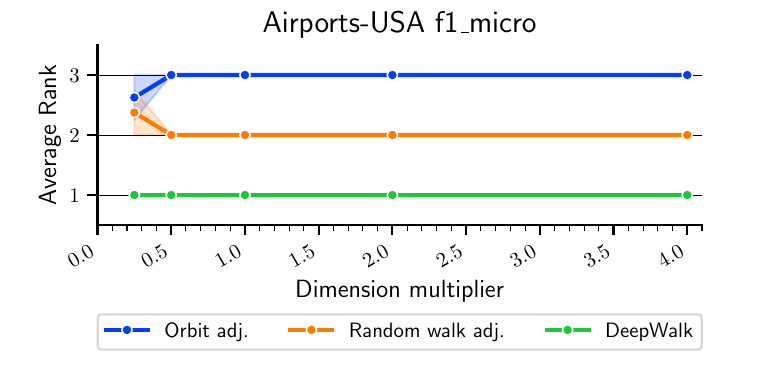} &
			\includegraphics[width=0.45\linewidth]{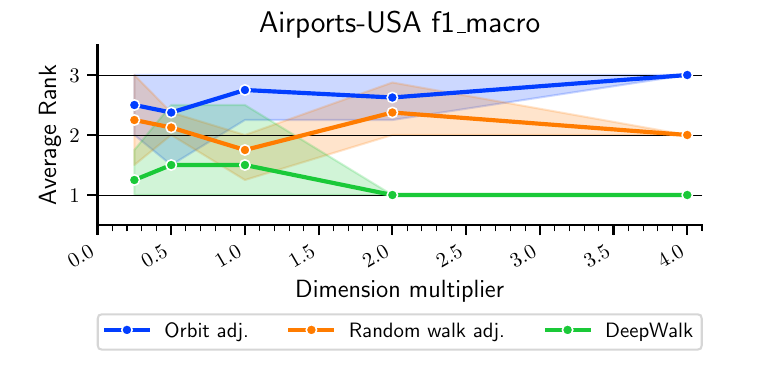} \\
			\includegraphics[width=0.45\linewidth]{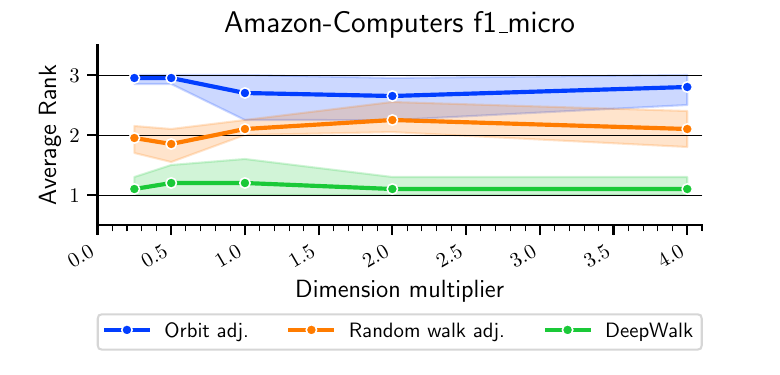} &
			\includegraphics[width=0.45\linewidth]{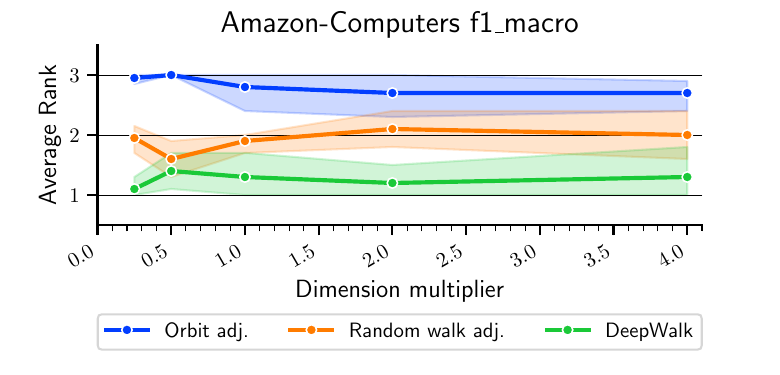} \\
			\includegraphics[width=0.45\linewidth]{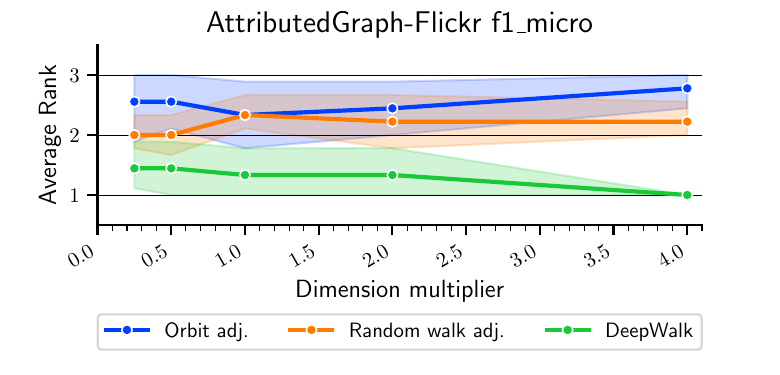} &
			\includegraphics[width=0.45\linewidth]{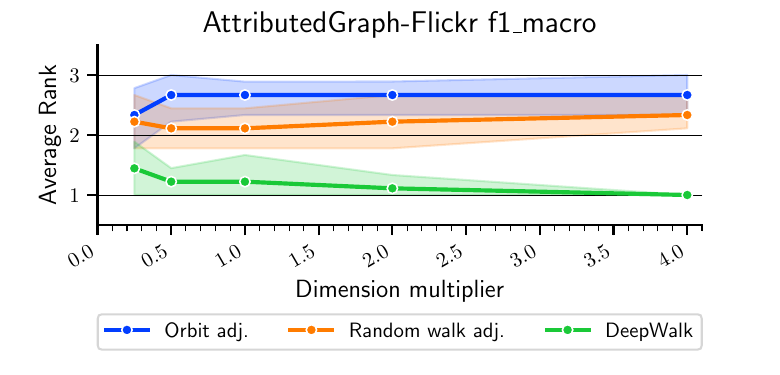} \\
			\includegraphics[width=0.45\linewidth]{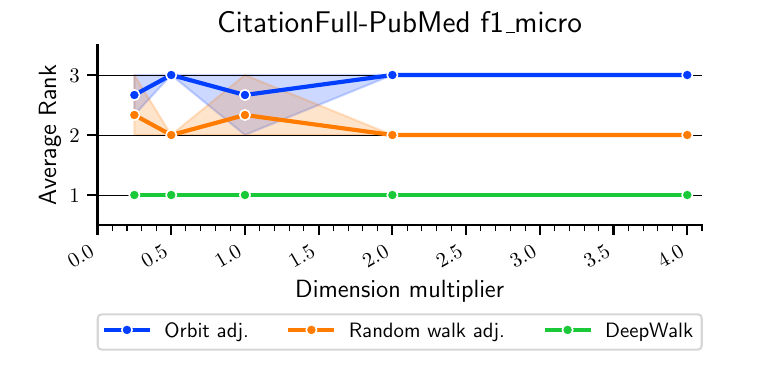} &
			\includegraphics[width=0.45\linewidth]{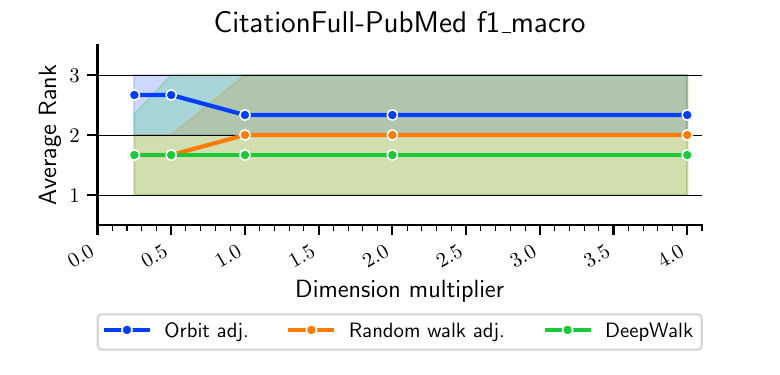} \\
			\includegraphics[width=0.45\linewidth]{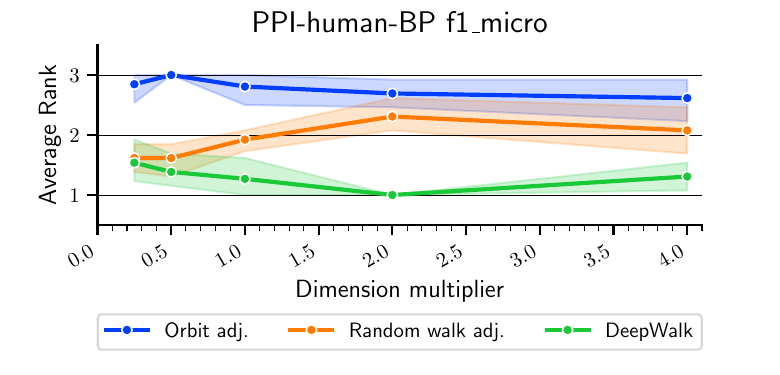} &
			\includegraphics[width=0.45\linewidth]{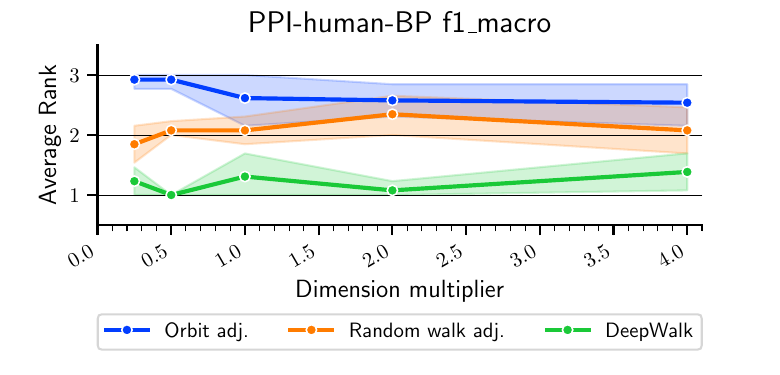} \\
			\includegraphics[width=0.45\linewidth]{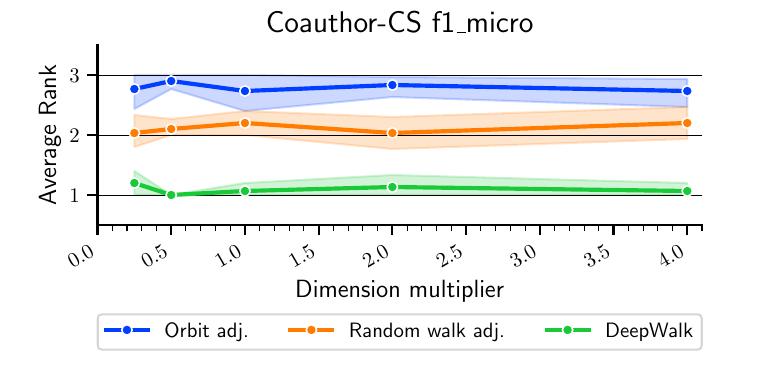} &
			\includegraphics[width=0.45\linewidth]{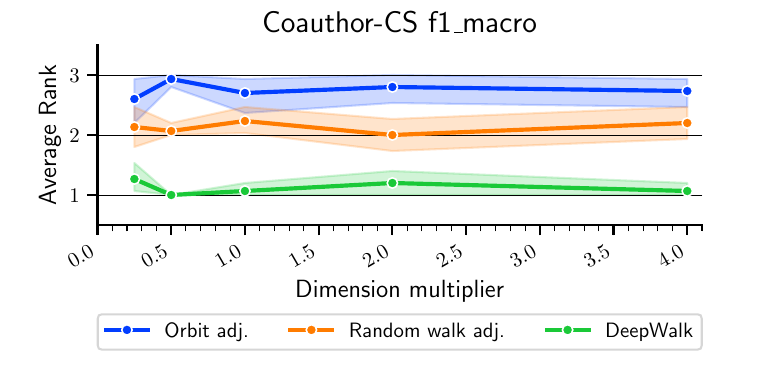} \\

		\end{tabular}
	\caption{{\bf Impact of the embedding space dimension on the ranking of the performance of the best type of adjacency across six different networks.}
			For each of our six networks (top to bottom), we vary the embedding space dimension from one quarter to four times the number of node classes (x-axis) and measure the average rank of the best performing adjacency for each embedding strategy (y-axis) using the micro averaged F1 score (left) and macro averaged F1 score (right). The error bands represent the 95\% confidence interval of the average rank, computed using bootstrapping.}
	\label{fig:dimension_sensitivity}
\end{figure}

\begin{figure}[h]
	\centering
	\begin{tabular}{cc}
	\includegraphics[width=0.45\linewidth]{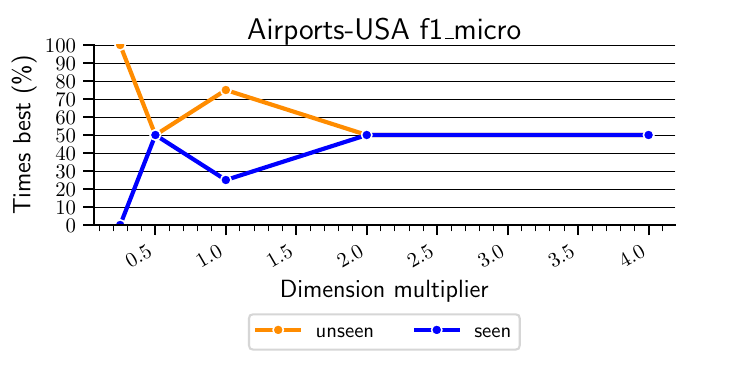} &
	\includegraphics[width=0.45\linewidth]{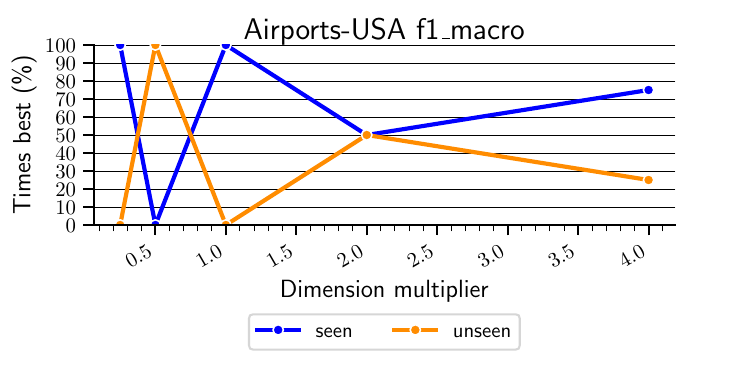} \\
	\includegraphics[width=0.45\linewidth]{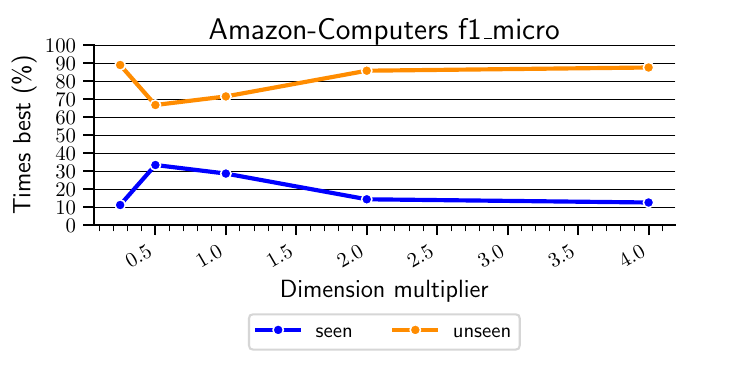} &
	\includegraphics[width=0.45\linewidth]{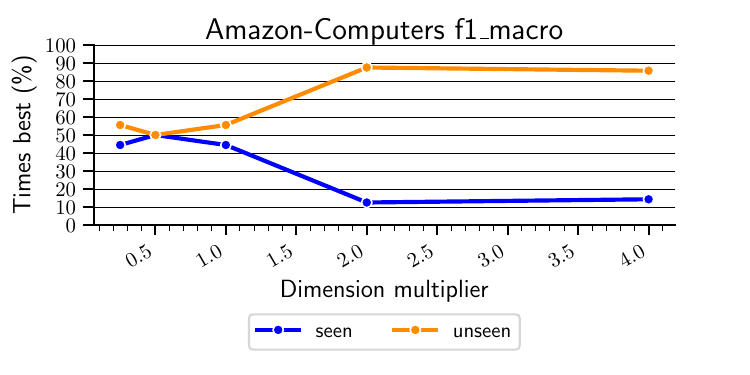} \\
	\includegraphics[width=0.45\linewidth]{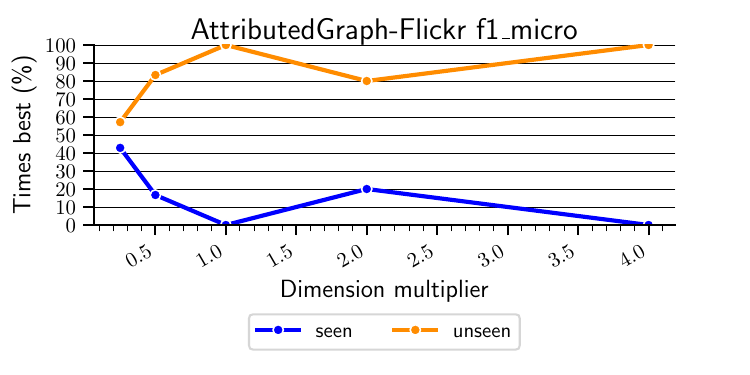} &
	\includegraphics[width=0.45\linewidth]{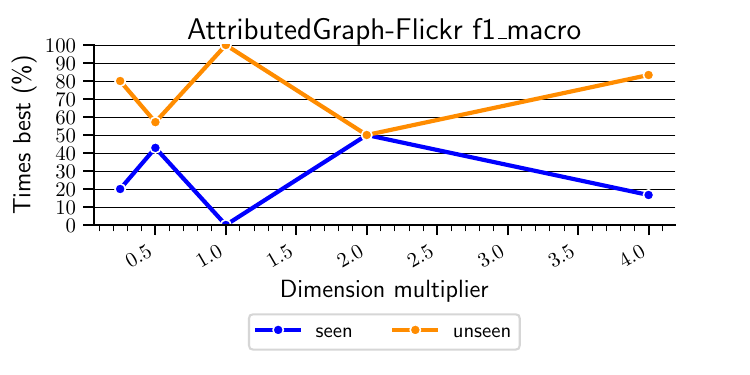} \\
	\includegraphics[width=0.45\linewidth]{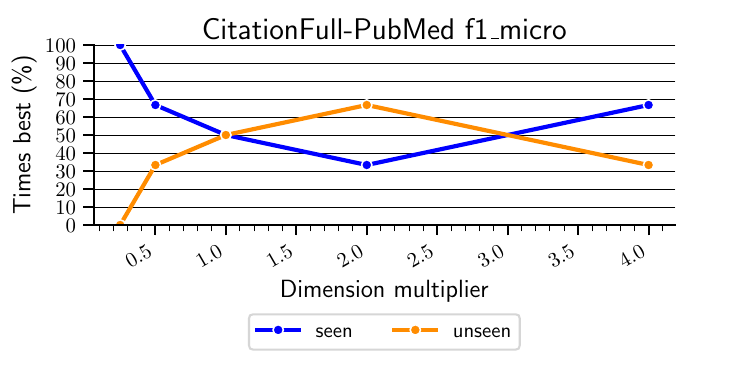} &
	\includegraphics[width=0.45\linewidth]{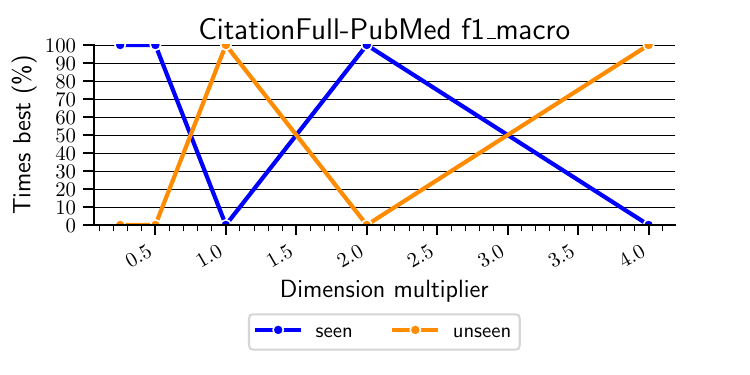} \\
	\includegraphics[width=0.45\linewidth]{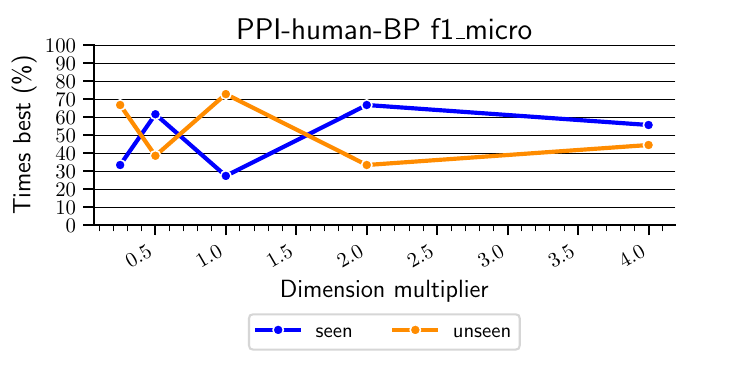} &
	\includegraphics[width=0.45\linewidth]{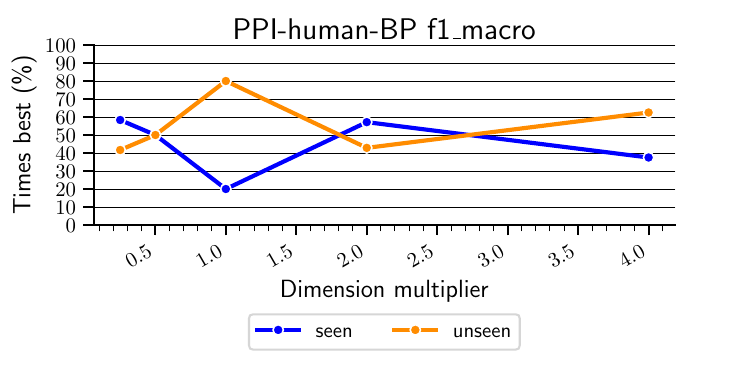} \\
	\includegraphics[width=0.45\linewidth]{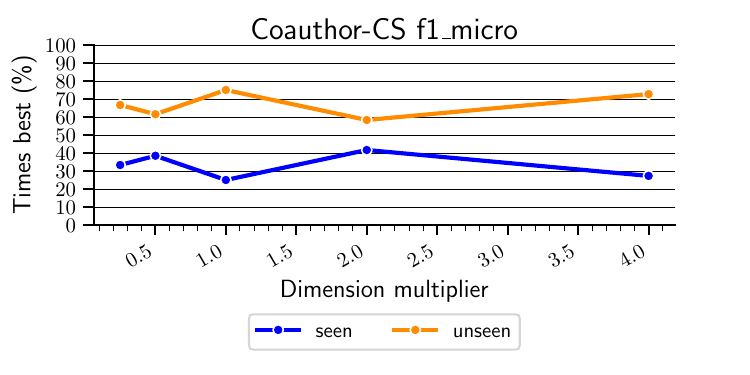} &
	\includegraphics[width=0.45\linewidth]{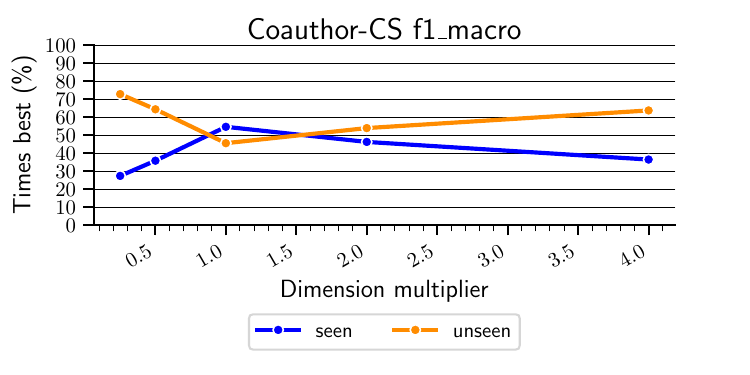} \\
	\end{tabular}
	\caption{{\bf The impact of the dimension of the embedding space on what proportion of the orbit adjacencies that outperform random walks adjacencies, are orbit adjacencies that are seen by random walks, across six different networks.} 
 For each of our six networks
	(top to bottom), we vary the embedding space dimension from one quarter
	to four times the number of node classes (x-axis) and show the
proportion of times the best performing orbit adjacency is unseen by random
walks up to length three (y-axis), when the orbit adjacency outperforms random
walk adjacency and DeepWalk, as measured using the micro averaged F1 score
(left) and the macro averaged F1 score(right).}
	\label{fig:dimension_impact_seen}
\end{figure}

\FloatBarrier

\subsection{Relation between orbit counts and orbit adjacency}
\label{sub:orbit_counts}

Orbit adjacency matrices count the number of times two nodes are orbit adjacent
with each node on a pre-specified given orbit. So, to get the number of times a
node occurs on a given orbit, we can simply sum over the corresponding row or
column in the orbit adjacency matrix. However, in some cases a node can touch
the same orbit adjacency multiple times whilst only occurring on a given orbit once.
For instance, when a node occurs on orbit adjacency \(\adjSingle{3}{3}\), i.e., is
part of a triangle, it occurs on orbit adjacency \(\adjSingle{3}{3}\) twice, once for each of
its neighbours, despite occurring on orbit 2 only once (it is part of the same triangle
only once). So, to correct for this, when computing the number of times a node
occurs on orbit 2 based on  \(\adjSingle{3}{3}\), we divide the orbit adjacency
counts by two. We illustrate this for the 14 orbits in the four-node graphlets
in Equation Set \ref{eqs:gdv_orbit_adjacency} below.

\begin{subequations}
\begin{align}
	o_1(x) &= \rowsum \adjDouble{1}{1}(x, y)  = \rowsum \adjSingle{1}{2}(x, y)      \\
	o_2(x) &= \rowsum \adjSingle{2}{1}(x, y)/2    \\
	o_3(x) &= \rowsum \adjSingle{3}{3}(x, y)/2      \\
	o_4(x) &= \rowsum \adjDouble{4}{4}(x, y)  = \rowsum \adjSingle{4}{5}(x, y) = \rowsum \adjDouble{4}{5}(x, y)      \\
	o_5(x) &= \rowsum \adjSingle{5}{5}(x, y)  = \rowsum \adjDouble{4}{5}(x, y)      \\
	o_6(x) &= \rowsum \adjDouble{6}{6}(x, y)  = \rowsum \adjSingle{6}{7}(x, y)      \\
	o_7(x) &= \rowsum \adjSingle{7}{6}(x, y)/3    \\
	o_8(x) &= \rowsum \adjSingle{8}{8}(x, y)/2  = \rowsum \adjDouble{8}{8}(x, y)      \\
	o_9(x) &= \rowsum \adjDouble{9}{10}(x, y)/2 = \rowsum \adjSingle{9}{11}(x, y)     \\
	o_{10}(x) &= \rowsum \adjDouble{10}{9}(x, y) = \rowsum \adjSingle{10}{11}(x, y)  \\
	o_{11}(x) &= \rowsum \adjDouble{11}{9}(x, y) = \rowsum \adjSingle{11}{10}(x, y) \\
	o_{12}(x) &= \rowsum \adjDouble{12}{12}(x, y) =  \rowsum \adjDouble{12}{13}(x, y)/2   \\
	o_{13}(x) &= \rowsum \adjDouble{13}{12}(x, y)/2 = \rowsum \adjDouble{13}{13}(x, y)   \\
	o_{14}(x) &= \rowsum \adjSingle{14}{14}(x, y)/3  
\end{align}
\captionof{equationset}{{\bf Computing orbit counts from orbit adjacency}}
\label{eqs:gdv_orbit_adjacency}
\end{subequations}

\subsection{Relation between graphlet adjacency and orbit adjacency}
\label{sub:graphlet_adjacency}

Orbit is a direct generalisation of graphlet adjacency, taking into account the
orbit positions of the nodes in the graphlet. So, to compute graphlet adjacency
from orbit adjacency for a given graphlet, we can simply sum the orbit adjacency matrices
for the orbits that are part of the graphlet. We illustrate this for the nine graphlets
adjacency matrices for up to four-node graphlets in Equation Set \ref{eqs:graphlet_orbit_adjacency} below.

\begin{subequations}
\begin{align}
	\gradj{1}  &= \adjDouble{1}{1} + \adjSingle{1}{2} + \adjSingle{2}{1} \\
	\gradj{2}  &= \adjSingle{3}{3}  \\
	\gradj{3}  &= \adjDouble{4}{4} + \adjSingle{4}{5} + \adjSingle{5}{4} + \adjDouble{4}{5} + \adjDouble{5}{5} + \adjSingle{5}{5} \\
	\gradj{4}  &= \adjDouble{6}{6} + \adjSingle{6}{7} + \adjSingle{7}{6} \\
	\gradj{5}  &= \adjSingle{8}{8} + \adjDouble{8}{8} \\
	\gradj{6}  &= \adjDouble{9}{10} + \adjDouble{10}{9} + \adjSingle{9}{11} + \adjSingle{11}{9} + \adjSingle{10}{10} + \adjSingle{10}{11} + \adjSingle{11}{10} \\
	\gradj{7}  &= \adjDouble{12}{12} + \adjSingle{12}{13} + \adjSingle{13}{12} + \adjSingle{13}{13} \\
	\gradj{8}  &= \adjSingle{14}{14} 
\end{align}
\captionof{equationset}{{\bf Computing graphlet adjacency from orbit adjacency}}
\label{eqs:graphlet_orbit_adjacency}
\end{subequations}

\subsection{GRaphlet-orbit ADjacency COunter (GRADCO)}
\label{sub:gradco}

To enable orbit adjacency analysis of real networks, we developed our
GRaphlet-orbit ADjacency COunter (GRADCO), which computes all 28 orbit
adjacencies for up to four-node graphlets.  To develop GRADCO, we used the
tried and tested strategy of state-the-art graphlet counters to compute some
type of orbit adjacency counts via the enumeration of specific, easy find
graphlets and to infer the remaining types of orbit adjacency counts from the
computed ones. GRADCO is implemented in C++ and is available as a Python
library via pip. 

In this section we first present our literature survey of the state-of-the art
graphlet and orbit counters (Section \ref{sub:counting_literature}).
Then, to enable our fast counter, we generalise the ORbit Counting Algorithm
(ORCA) system of orbit redundancy equations to determine two sets of equations
that determine linear relations between orbit adjacencies for orbits that are one and two hops apart
(Section \ref{sub:orbit_adjacency_equations}). To make these two systems determined, we implement Grochow’s strategy to quickly
enumerate all cliques (graphlet \(G_8\)) and chordal cycles (graphlet \(G7\)) to compute 
orbit adjacencies \(\adjSingle{14}{14}\) and \(\adjDouble{12}{12}\) (Section \ref{sub:brute_force_orbit_adjacency}). Then, as orbit
adjacency \(\adjTriple{4}{4}\) is not covered by our equations and brute force
enumeration of all four-node paths is very time consuming, we present our
solution to compute \(\adjTriple{4}{4}\) based on simple and more efficient matrix operations. We 
use the fact that \(A^3\) can be seen as the sum of orbit
adjacencies, which include \(\adjTriple{4}{4}\), and simply subtract the 
orbit adjacencies that are not \(\adjTriple{4}{4}\) from the sum of all orbit
adjacencies that comprise \(A^3)\) (see Section \ref{sub:compute_adjTriple}). 
 Finally, to illustrate the expected time performance of GRADCO, we measure
 GRADCO's running time on well-studied model networks (see Section
 \ref{sub:time_performance}). 

\subsubsection{Graphlet counting literature}
\label{sub:counting_literature}

The first algorithms to compute graphlets and orbit counts for up to \(k\)-node
graphlets do so by exhaustively enumerating all \(k\)-node subgraphs in the
network and counting the encountered graphlets (orbits) along the way
\citep{Przulj_2004_modeling, prvzulj2007biological}. These counters enumerate all
\(k\)-node subgraphs in a given network by exploring each node's neighbourhood
up to \(k-1\)-hops deep using \emph{depth-first search} (DFS). For
a given starting node, DFS recursively `hops' away from the starting node,
visiting one of the starting node's direct neighbours and from there, move on
to one of that node's neighbours and so on, until \(k-1\) hops have been made.
Then, for the subgraph induced by the \(k\) nodes
visited (i.e., including the start node), the occurring graphlets are
registered, before backtracking and continuing the DFS. The DFS continues until
all nodes in the network have been considered as a starting node. Hence, say
that the maximum number of neighbours of a node in the network, also known as
a node's \emph{degree}, is \(d\), then the time complexity to count \(k\)-node graphlets
using DFS is \(\mathcal{O}(nd^{k-1})\). This is problematic as many real networks
contain many very high degree nodes, as their degree distribution often follows
a power law: \(P(d) \sim d^{-\lambda}\), where \(\lambda\) usually ranges between 2
and 3 \citep{Ravasz_2003}. Additionally, using this naive approach, the same
sets of nodes are sometimes visited multiple times in a different order,
leading to some graphlets being
\emph{overcounted}. For instance, when three-nodes form a triangle in the
network, DFS visits those same three-nodes six times, but each time in a different
ordering. 

Various strategies have been proposed to speed up graphlet (orbit) counting.
Implicitly, they all aim to limit the search space by limiting the number of
times the same sets of nodes are visited during graph traversal.  The strategy
used in state-of-the-art counters was first proposed by
Grochow \cite{Grochow_2007_network}.  To count the occurrences of a given graphlet
\(G_k\) in an undirected network \(H=\{V, E\}\), they first convert \(H\) into a
directed acyclic graph, \(\gdag=\{V, \edag\}\), by orienting all edges
in \(\edag\) so that each edge goes from the node with lower degree in \(H\) to the
node with higher degree in \(H\). We denote this degree ordering of \(H\) as
\(\prec\), where node \(u \prec v\) if \(d(u) < d(v)\). Then, they illustrate that
for real networks, the maximum number of outgoing edges of a node in \(\gdag\),
\(o\), is usually a lot less than \(d\), i.e., \(o \ll d\).  Additionally,
they posit that to count the occurrences of a given graphlet \(G_k\) in the
network \(H\), it suffices to count the different non-isomorphic directed orientations of that
graphlet in \(\gdag\).
So, combining both ideas, to count the
occurrences of a given graphlet \(G_k\) in \(H\), they count the different directed
orientations of that graphlet in \(\gdag\) instead, traversing \(\gdag\) over
out-edges instead of in-edges as much as possible. Applying this strategy to
count four-node cliques, for instance, has a time complexity of \(\bigO{
no^3} \ll \bigO{nd^3}\). Moreover, they avoid visiting
isomorphisms of the same subgraph multiple times completely by applying a
constraint to the ordering of the nodes that occur on the same symmetric
position (i.e., orbits) within an enumerated subgraph. For instance, when
traversing \(\gdag\) to enumerate all directed three-node paths with two edges
pointing out from the centre node, e.g., a path \(a \leftarrow b \rightarrow c\),
then they require for the peripheral nodes \(a\) and \(c\), that \(c\) precedes \(a\) in
the degree ordering to avoid visiting this subgraphs twice (e.g., as \(a
\leftarrow b \rightarrow c\) and \(c \leftarrow b \rightarrow a\)).

\subsubsection{Orbit adjacency redundancy equations for 4-node graphlets}
\label{sub:orbit_adjacency_equations}

We generalise ORCA's redundancy equations to determine the linear
relations between the different orbit adjacency counts. We define two systems
of equations: ten equations for orbit adjacencies in which the two orbits are
one hop away and four equations for orbit adjacencies in which the two orbits
are two hops apart (see Equation Set \ref{eqs:orbit_equations_one} and
\ref{eqs:orbit_equations_two}, respectively). We explain the general
idea behind these equations and provide an example by building our first
single-hop equation that relates \(\adjSingle{12}{13}(x, y)\) +
\(\adjSingle{14}{14}(x, y)\).

The ORCA redundancy equations are based on the following idea. First, we
pick a three-node graphlet with nodes \(x, y \text{ and } z\) on predefined orbits.
Then, we create a four-node graphlet by adding a fourth node, \(w\), according to
one of two strategies: \(w\) has to either form a path or a triangle with two of
the three-nodes \(x, y \text{ and } z\). Adding \(w\) using either of those strategies, we
know for seven of the eight possible edges between \(x, y, z\) and \(w\) whether
they are present or absent. Hence, \(x, y, z \text{ and } w\) are known to form
one of two four-node graphlets. Or, phrased at the orbit level, the four-nodes
are known to be on one of two orbits. Similarly, at the orbit adjacency level,
it is known for each of the pairs of nodes \((x, y), (x, z), (x, w), (y, z), (y,
w) \text{ and } (z, w)\) that they are orbit adjacent with respect to one of two
orbit adjacencies. However, setting up the systems of equations
this way would be too expensive: enumerating a given three-node graphlet,
extending it with all possible `fourth' nodes (i.e., all possible nodes \(w_i\)),
each time checking if that node extension forms a path or triangle with two of
the three-nodes, would be as expensive as enumerating all four-node graphlets.
So, instead, we pre-compute for all pairs of nodes their frequency of forming a
path and a triangle together. This is equivalent to computing and storing orbit adjacency
matrices \(\adjSingle{1}{2}\) and \(\adjSingle{3}{3}\). Then, instead of adding one
node \(w\) at a time, we can look up for two of three-nodes in \(x, y \text{ and }
z\), their frequency of co-occurrence on a triangle or a path. Setting up the system
of equations this way has the same time complexity of enumerating all three
node subgraphs. 

\begin{figure}[H]
	\centering
	\includegraphics[width=.99\columnwidth]{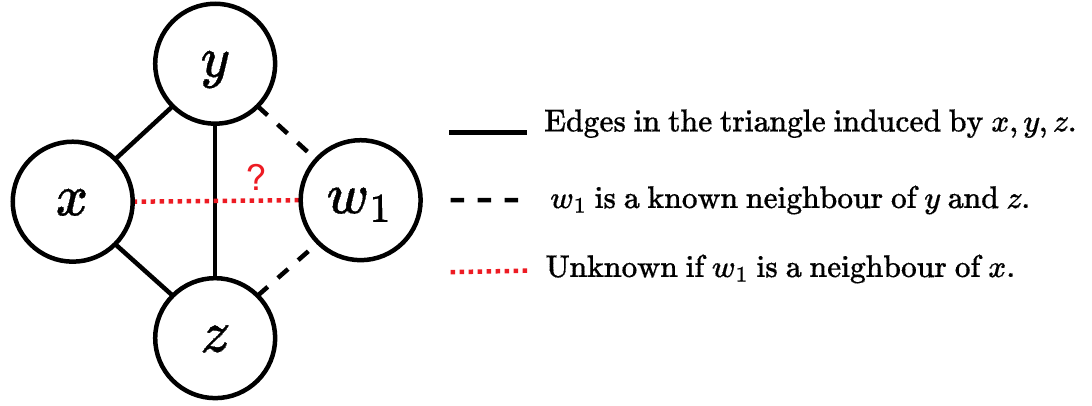}
	\caption{{\bf Relation between orbit adjacency \(\adjSingle{14}{14}\) and
	\(\adjSingle{12}{13}\).} Solid lines are edges in the three-node
graphlet. Dashed lines are edges to the three common neighbours of \(y\) and \(z\):
\(w_{i \in {1,2,3}}\). Dotted red lines are unknowns: their presence or absence
makes the resulting four-node graphlet on \(x, y, z \text{ and } w_i\) isomorphic
to either \(G_7\) or \(G_8\). That is, depending on whether \(w_i\) is connected to \(x\) or not, \(x\) and \(y\) are orbit adjacent with
respect to \(\adjSingle{14}{14}\) or \(\adjSingle{12}{13}\), respectively.}
\label{fig:orbit_adjacency_clique} \end{figure}

We illustrate our first single-hop equation, which relates
\(\adjSingle{12}{13}(x, y)\) and \(\adjSingle{14}{14}(x, y)\), in
\fref{fig:orbit_adjacency_clique}. Nodes \(\xyz\) induce graphlet \(G_2\) (a
triangle). There are three-nodes: \(w_1, w_2 \text{ and } w_3\) that are
connected to both \(y\) and \(z\) (i.e. \(\adjSingle{3}{3}(y,z)=4\), including the
triangle with \(x\)). For each node \(w_{i \in {1,2,3}}\) that is connected to \(x\),
nodes \(x, y, z \text{ and } w_i\) form a four-node clique (i.e., graphlet \(G_8\))
and nodes \(x\) and \(y\) are orbit adjacent with respect to \(\adjSingle{14}{14}\).
For each node \(w_{i \in {1,2,3}}\) that is not connected to \(x\), nodes \(x, y, z
\text{ and } w_i\) form a cycle with a single chord (i.e., graphlet \(G_7\)) and
nodes \(x\) and \(y\) orbit adjacent with respect to \(\adjSingle{12}{13}\). Applying
an ordering on \(H\), we sum over all unique triangles \(G[{x, y, z}] \cong G_2\)
in \(H\) (i.e., avoiding visiting multiple isomorphisms of the same triangle
\(G_2\)). We also take into account symmetries: each node \(w_i\) that is connected
to \(x\) could also occur on the position of \(x\) and \(y\). Accounting for this we
get: 
\begin{equation} 
	\adjSingle{12}{13}(x, y) + 2 \adjSingle{14}{14}(x, y) =
\gitertwo  \adjSingle{3}{3}(y,z) -1 
\end{equation}

In practice, there are ten non-redundant ways to combine the enumeration of a
given type of three-node graphlet with one of the two extension strategies to
get the combined counts of two single-hop orbit adjacencies and four ways to
get the combined counts of two double-hop orbit adjacencies, determining two
sets of equations (see Equation Sets \ref{eqs:orbit_equations_one}
and \ref{eqs:orbit_equations_two}, respectively). Then, by computing one
type of four-node graphlet, orbit, or orbit adjacency count through exhaustive
enumeration, the system of equations is determined and the remaining counts can
be inferred. 


\begin{subequations}
\begin{align}
	\adjSingle{12}{13}(x, y) + 2 \adjSingle{14}{14}(x, y) &= \gitertwo  \adjSingle{3}{3}(y,z) -1 \\
	2 \adjSingle{13}{13}(x, y) + 2 \adjSingle{14}{14}(x, y) &= \gitertwo  \adjSingle{3}{3}(x,y)  -1 \\
	\adjSingle{10}{10}(x, y) + \adjSingle{12}{13}(x, y) &= \gitertwo  \adjSingle{1}{2}(y,z)    \\
	\adjSingle{10}{11}(x, y) + \adjSingle{12}{13}(x, y) &= \gitertwo  \adjSingle{1}{2}(y,x)    \\
	2 \adjSingle{6}{7}(x, y)   + \adjSingle{10}{11}(x, y) &= \giteroneA \adjSingle{1}{2}(y,x) -1 \\
	\adjSingle{5}{5}(x, y)   + \adjSingle{8}{8}(x, y)   &= \giteroneA \adjSingle{1}{2}(x,y)    \\
	2 \adjSingle{6}{7}(x, y)   + 2 \adjSingle{9}{11}(x, y)  &= \giteroneA \adjSingle{1}{2}(x,y) -1 \\
	2 \adjSingle{9}{11}(x, y)  + \adjSingle{12}{13}(x, y) &= \giteroneB \adjSingle{3}{3}(y,z)    \\
	\adjSingle{4}{5}(x, y)   + \adjSingle{8}{8}(x, y)   &= \giteroneB \adjSingle{1}{2}(y,z)    \\
	\adjSingle{8}{8}(x, y)   + \adjSingle{12}{13}(x, y) &= \giteroneB \adjSingle{1}{2}(y,z) -1 
\end{align}
\captionof{equationset}{{\bf Orbit-adjaceny redundancy equations for directly connected nodes}}
\label{eqs:orbit_equations_one}
\end{subequations}

\begin{subequations}
\begin{align}
	\adjDouble{6}{6}(x, y)   + \adjDouble{9}{10}(x, y)  &= \giteroneB \adjSingle{3}{3}(y,z)    \\ 
	\adjDouble{9}{10}(x, y)  + 2\adjDouble{12}{12}(x, y) &= \giteroneB \adjSingle{3}{3}(y,z)    \\ 
	\adjDouble{4}{5}(x, y)   + 2\adjDouble{8}{8}(x, y)   &= \giteroneB \adjSingle{1}{2}(y,z) -1 \\ 
	2 \adjDouble{8}{8}(x, y)   + 2\adjDouble{12}{12}(x, y) &= \giteroneB \adjSingle{1}{2}(y,z) -1  
\end{align}
\captionof{equationset}{{\bf Orbit-adjaceny redundancy equations for nodes two hops appart.}}
\label{eqs:orbit_equations_two}
\end{subequations}

\FloatBarrier

\subsubsection{Brute-force orbit adjacency counting}
\label{sub:brute_force_orbit_adjacency}

To solve our two system of equations for the single-hop and double-hop orbit
adjacencies, we need to compute one single-hop and one double-hop orbit
adjacency matrix for each system (see Section \ref{sub:orbit_adjacency_equations}). 
To efficiently enumerate a given graphlet Grochow et al. propose to apply a degree
ordering to the graph and to count the different non-isomorphic directed
orientations of that graphlet in the directed acyclic graph thus achieved
\cite{Grochow_2007_network}.  We use this strategy to efficiently enumerate all
four node cliques (graphlet \(G_8\)) and chordal cycles (graphlet \(G7\)) to
compute orbit adjacencies \(\adjSingle{14}{14}\) and \(\adjDouble{12}{12}\),
respectively \cite{Grochow_2007_network}, making our two systems of equations
determined.  We also use the strategy of Grochow et al. to enumerate all
three-node paths and triangles to compute all three node graphlet based orbit
adjacencies, \(\adjSingle{1}{2}\), \(\adjDouble{1}{1}\) and
\(\adjSingle{3}{3}\), and to set up our system of equations.

\begin{figure}[h]
	\centering
	\includegraphics[width=0.50\linewidth]{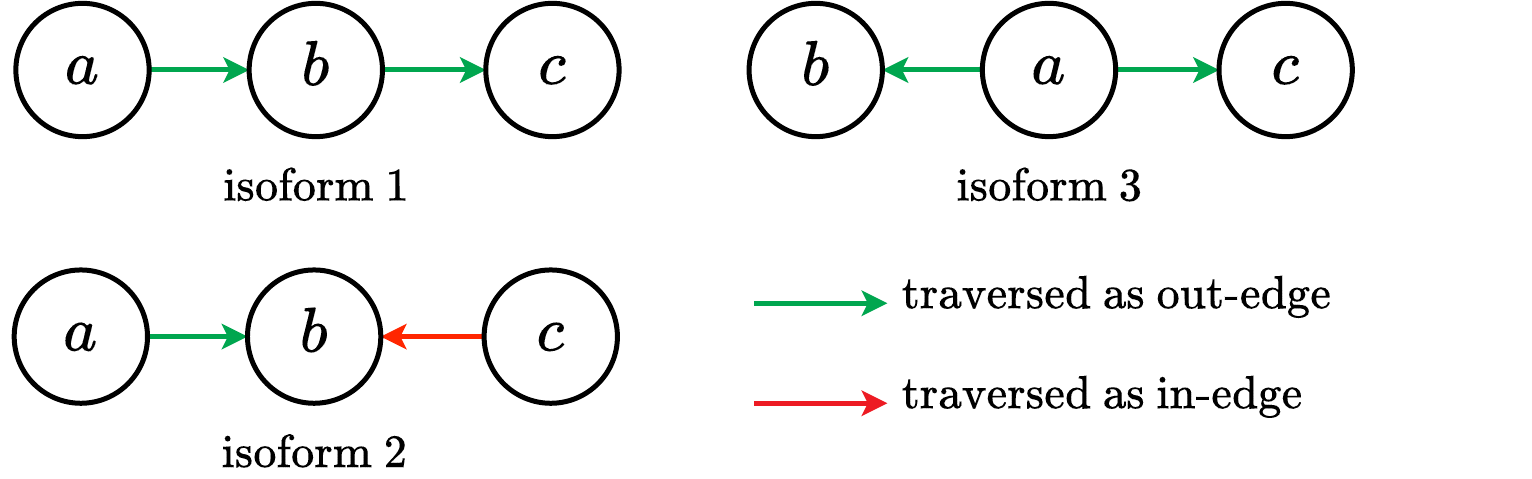}
	\caption{Isomorphic directed orientations of a three-node path (graphlet \(G_1\)) and how they are enumerated.}
	\label{fig:enumerate_G1}
\end{figure}

\begin{figure}[h]
	\centering
	\includegraphics[width=0.50\linewidth]{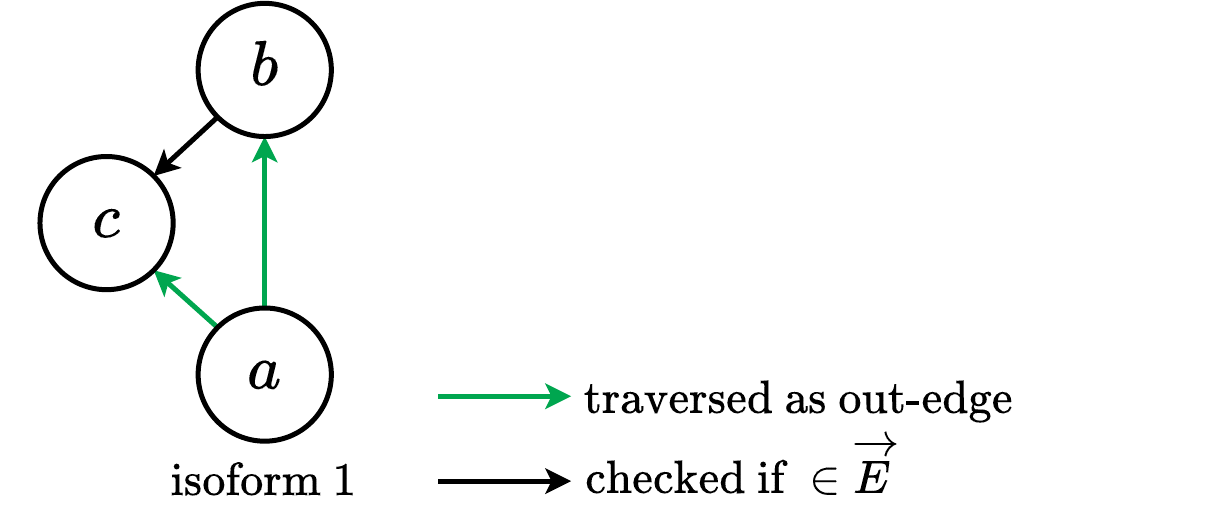}
	\caption{Isomorphic directed orientations of a triangle (graphlet \(G_2\)) and how they are enumerated.}
	\label{fig:enumerate_G2}
\end{figure}

\begin{figure}[h]
	\centering
	\includegraphics[width=0.75\linewidth]{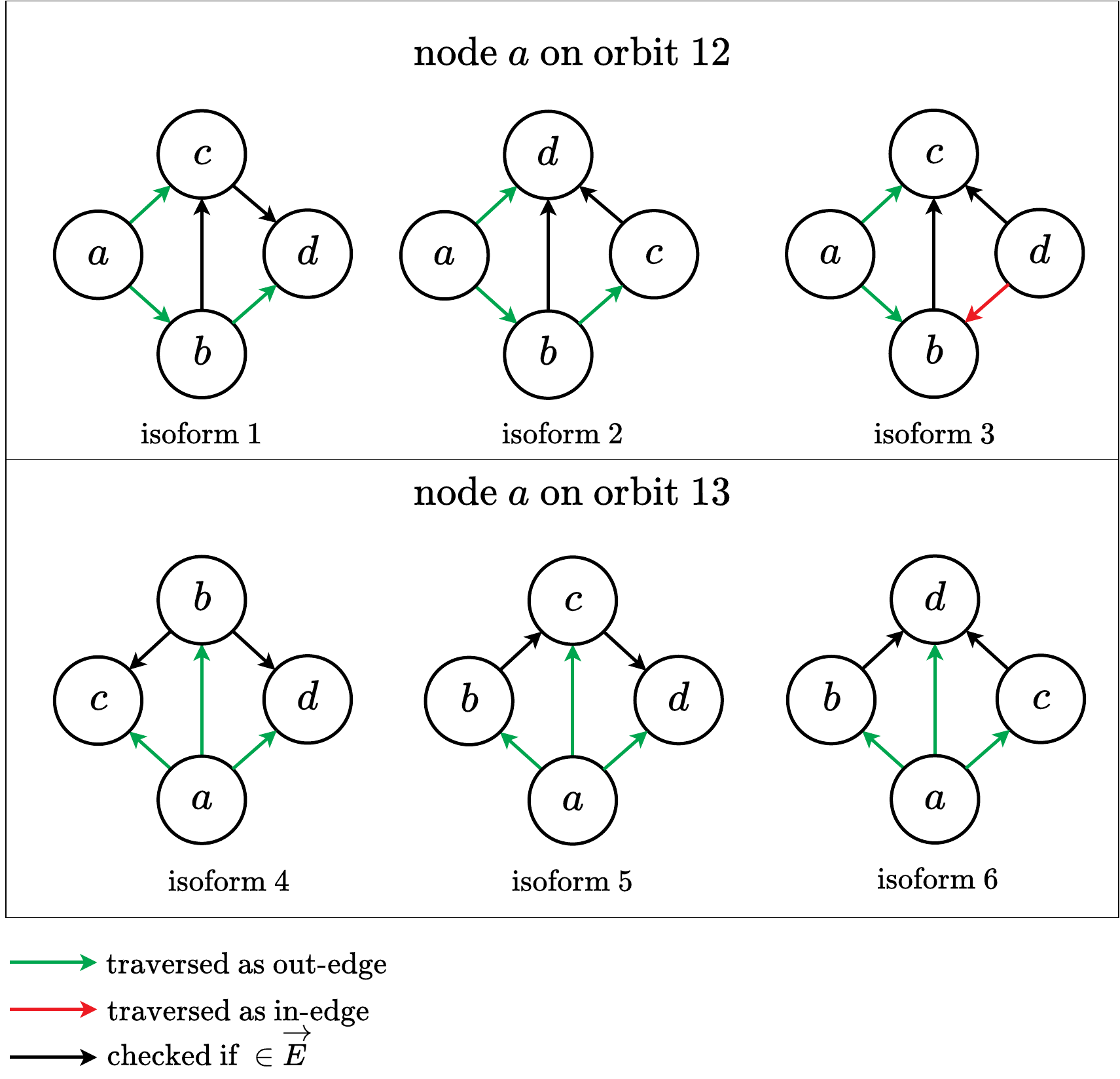}
	\caption{Isomorphic directed orientations of a chordal cycle (graphlet \(G_7\)) and how they are enumerated.}
	\label{fig:enumerate_G7}
\end{figure}

\begin{figure}[h]
	\centering
	\includegraphics[width=0.50\linewidth]{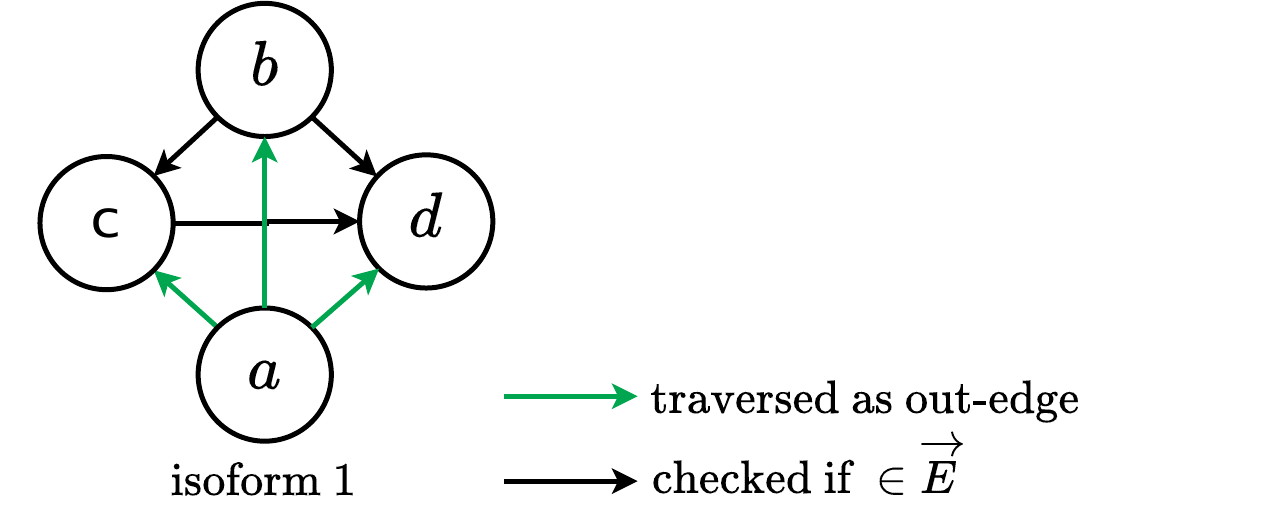}
	\caption{Isomorphic directed orientations of a clique (graphlet \(G_8\)) and how they are enumerated.}
	\label{fig:enumerate_G8}
\end{figure}

In figures \ref{fig:enumerate_G1} to \ref{fig:enumerate_G8} we illustrate the
different isomorphic directed orientations of a three-node path, a triangle, a
chordal cycle and a clique, respectively, and how we enumerate them.  In
algorithms \ref{alg:enumerate_G1} to \ref{alg:enumerate_G8} we provide the
pseudo code to enumerate the different isomorphic directed orientations of a
three-node path, a triangle, a chordal cycle and a clique, respectively.

\FloatBarrier

\begin{algorithm}[]
\caption{Counting \(\adjSingle{1}{2}\) and \(\adjDouble{1}{1}\) via the exhaustive enumeration of the directed orientations of a three-node path (\(G_1\)).}
\KwIn{An unweighted undirected graph \(H=(V, E)\).}
\KwOut{The orbit adjacency matrices \(\adjSingle{1}{2}\) and \(\adjDouble{1}{1}\).}
\BlankLine
Order de nodes in \(H\) according to their degree\\
Apply the ordering to \(H\) to obtain \(\gdagve\)\\
\For{\(a \in V\)}{
	\For{\( \text{each out-neighbour } b \text{ of } a\)}{
		\For{\( \text{each out-neighbour } c \text{ of } b\)}{
			\If{ \(e(a, c) \notin E\)}{
			\tcp{\(G_1\) isoform 1: \(a \rightarrow b \rightarrow c\)}
			\(\adjSingle{1}{2}(a, b) \texttt{++}\)\\
			\(\adjSingle{1}{2}(c, a) \texttt{++}\)\\
			\(\adjDouble{1}{1}(a, c) \texttt{++}\)\\
			\(\adjDouble{1}{1}(c, a) \texttt{++}\)\\
			}
		}
	\For{\( \text{each in-neighbour } c \text{ of } b\text{, with }a \prec c\)}{
			\If{\(e(a, c) \notin \ed\)}{
			\tcp{\(G_1\) isoform 2: \(a \rightarrow b \leftarrow c\)}
			\(\adjSingle{1}{2}(a, b) \texttt{++}\)\\
			\(\adjSingle{1}{2}(c, a) \texttt{++}\)\\
			\(\adjDouble{1}{1}(a, c) \texttt{++}\)\\
			\(\adjDouble{1}{1}(c, a) \texttt{++}\)\\
			}
		}
	\For{\( \text{each out-neighbour } c \text{ of } a \text{, with }b \prec c\)}{
			\If{\( e(b, c) \notin \ed \)}{
			\tcp{\(G_1\) isoform 3:  \(b \leftarrow a \rightarrow c\)}
			\(\adjSingle{1}{2}(b, a) \texttt{++}\)\\
			\(\adjSingle{1}{2}(c, a) \texttt{++}\)\\
			\(\adjDouble{1}{1}(b, c) \texttt{++}\)\\
			\(\adjDouble{1}{1}(c, b) \texttt{++}\)\\
			}
		}
	}
}
\label{alg:enumerate_G1}
\end{algorithm}

\begin{algorithm}[]
\caption{Counting \(\adjSingle{3}{3}\) via the exhaustive enumeration of the directed orientations of a triangle (\(G_2\)).}
\KwIn{An unweighted undirected graph \(H=(V, E)\).}
\KwOut{The orbit adjacency matrix \(\adjSingle{3}{3}\).}
\BlankLine
Order de nodes in \(H\) according to their degree\\
Apply the ordering to \(H\) to obtain \(\gdagve\)\\
\For{\(a \in V\)}{
	\For{\( \text{each out-neighbour } b \text{ of } a\)}{
		\For{\( \text{each in-neighbour } c \text{ of } b \text{, larger than } a\)}{
			\If{ \(e(a, c) \in \ed \)}{
			\tcp{\(G_2\) isoform 1}
			\(\adjSingle{3}{3}(a, b) \texttt{++}\)\\
			\(\adjSingle{3}{3}(b, a) \texttt{++}\)\\
			\(\adjSingle{3}{3}(a, c) \texttt{++}\)\\
			\(\adjSingle{3}{3}(c, a) \texttt{++}\)\\
			\(\adjSingle{3}{3}(b, c) \texttt{++}\)\\
			\(\adjSingle{3}{3}(c, b) \texttt{++}\)\\
			}
		}
	}
}
\label{alg:enumerate_G2}
\end{algorithm}

\begin{algorithm}[]
\caption{Counting \(\adjDouble{12}{12}\) via exhaustive enumeration enumeration of the directed orientations of a chordal cycle (\(G_7\)).}
\KwIn{An unweighted undirected graph \(H=(V, E)\).}
\KwOut{The orbit adjacency matrix \(\adjDouble{12}{12}\).}
\BlankLine
Order de nodes in \(H\) according to their degree\\
Apply the ordering to \(H\) to obtain \(\gdagve\)\\
\For{\(a \in V\)}{
	\For{\( \text{each out-neighbour } b \text{ of } a\)}{
		\For{\( \text{each out-neighbour} c \text{ of } a \text{, with } b \prec c\)}{
			\tcp{cases where node \(a\) might be on orbit 12}
			\If{\(e(b,c) \in E\)}{
				\For{\( \text{each out-neighbour } d \text { of } b\)}{
					\If{ \(e(a, d) \notin \ed\) and (\(e(c, d) \in \ed\) or \(e(d, c) \in \ed\)) }{
					\tcp{\(G_7\) isoform 1 or 2 }
					\(\adjDouble{12}{12}(a, d)\)++\\
					\(\adjDouble{12}{12}(d, a)\)++\\
					}
				}
				\For{\( \text{each in-neighbour } d \text { of } b \text{, with } a \prec d\)}{
					\If{ \(e(a, d) \notin \ed\) and \(e(d, c) \in \ed\) }{
					\tcp{\(G_7\) isoform 3}
					\(\adjDouble{12}{12}(a, d)\)++\\
					\(\adjDouble{12}{12}(d, a)\)++\\
					}
				}
			
			}
			\tcp{cases where node \(a\) might be on orbit 13}
			\For{\( \text{each out-neighbour } d \text{ of } a \text{, with } c \prec d\)}{
				\If{ \(e(b, c) \in \ed\) and \(e(b, d) \in \ed\)  and \(e(c, d) \notin \ed\)}{
				\tcp{\(G_7\) isoform 4}
					\(\adjDouble{12}{12}(c, d)\)++\\
					\(\adjDouble{12}{12}(d, c)\)++\\
				}
				\If{ \(e(b, c) \in \ed\) and \(e(b, d) \notin \ed\) and \(e(c, d) \in \ed\)}{
				\tcp{\(_7\) isoform 5}
					\(\adjDouble{12}{12}(b, d)\)++\\
					\(\adjDouble{12}{12}(d, b)\)++\\
				}
				\If{ \(e(b, c) \notin \ed\) and \(e(b, d) \in \ed\) and \(e(c, d) \in \ed\)}{
				\tcp{\(G_7\) isoform 6}
					\(\adjDouble{12}{12}(b, c)\)++\\
					\(\adjDouble{12}{12}(c, b)\)++\\
				}
			}
		}
	}
}
\label{alg:enumerate_G7}
\end{algorithm}

\begin{algorithm}[]
\caption{Counting \(\adjSingle{14}{14}\) via exhaustive enumeration}
\KwIn{An unweighted undirected graph \(H=(V, E)\)}
\KwOut{The orbit adjacency matrix \(\adjSingle{14}{14}\).}
\BlankLine
Order de nodes in \(H\) according to their degree\\
Apply the ordering to \(H\) to obtain \(\gdagve\)\\
\For{\(a \in V\)}{
	\For{\( \text{each out-neighbour } b \text{ of } a\)}{
		\For{\( \text{each out-neighbour } c \text{ of } b\)}{
			\If{ \(e(a, c) in \ed \)}{
			\tcp{\(G_2\) isoform 1}
			\For{\( \text{each out-neighbour } d \text{ of } c\)}{
				\If{ \(e(a, d) in \ed \) and \(e(b, d) in \ed \)}{
					\tcp{\(G_8\) isoform 1}
					\(\adjSingle{14}{14}(a, b) \texttt{++}\)\\
					\(\adjSingle{14}{14}(a, c) \texttt{++}\)\\
					\(\adjSingle{14}{14}(a, d) \texttt{++}\)\\
					\(\adjSingle{14}{14}(b, a) \texttt{++}\)\\
					\(\adjSingle{14}{14}(b, c) \texttt{++}\)\\
					\(\adjSingle{14}{14}(b, d) \texttt{++}\)\\
					\(\adjSingle{14}{14}(c, a) \texttt{++}\)\\
					\(\adjSingle{14}{14}(c, b) \texttt{++}\)\\
					\(\adjSingle{14}{14}(c, d) \texttt{++}\)\\
					\(\adjSingle{14}{14}(d, a) \texttt{++}\)\\
					\(\adjSingle{14}{14}(d, b) \texttt{++}\)\\
					\(\adjSingle{14}{14}(d, c) \texttt{++}\)\\
					}
				}
			}
		}
	}
}
\label{alg:enumerate_G8}
\end{algorithm}

\FloatBarrier

\subsubsection{Computing orbit adjacency \(\adjTriple{4}{4}\) via matrix operations}
\label{sub:compute_adjTriple}

As our redundancy equations do not cover \(\adjTriple{4}{4}\), we need a
different strategy to compute it. However, as real networks tend to be sparse,
four node paths occur very frequently in real-networks computing them via
exhaustive enumeration is very slow, even when applying a degree ordering
strategy. 

So, instead, we solve equation \ref{eq:random_walks_3}, which expresses random
walk adjacency matrix $A^3$ as the sum of orbit adjacency matrices, for
\(\adjTriple{4}{4}\):
\begin{equation}
\begin{split}
\adjTriple{4}{4} & =
A^3 - 
\adjSingle{0}{0} 
- \adjSingle{1}{2}
- \adjSingle{2}{1}
- 2 \adjSingle{3}{3} 
- \adjTriple{4}{4}
- \adjSingle{8}{8} \\
& \phantom{ = } - 2 \adjDouble{9}{10}
- 2 \adjDouble{10}{9}
- \adjDouble{12}{12} 
- \adjSingle{12}{13}
- \adjSingle{13}{12}
- \adjSingle{14}{14}
\end{split}
\end{equation}
As this requires the computation of $A^3$, and thus
requires the multiplication of the adjacency matrix, the time complexity of
this operation is \(\bigO{n^3}\).

\subsubsection{Time performance benchmarking}
\label{sub:time_performance}

\begin{figure}[h]
	\centering
	\includegraphics[width=0.8\columnwidth]{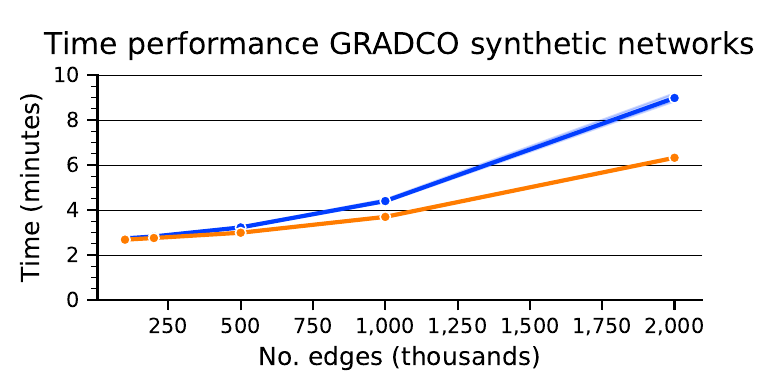}
  \caption{Running times of GRADCO for Erd\H{o}s-R\'enyi (ER) and Barab\'asi-Albert
Scale-Free (SF) model networks. For each model, the number of 
nodes is fixed to 20,000. The number of edges ranges between 100,000 and 2,000,000 (x-axis).
The error bands represent the 95\% confidence interval of the mean running time using bootstrapping.}
	\label{fig:times}
\end{figure}

To illustrate the expected time performance of GRADCO we measure GRADCO's
running time on well-studied model networks. Two factors influence the running
time of GRADCO: the enumeration of four-node cliques and chords,
\(\bigO{no^2i}\), and computing \(\adjTriple{4}{4}\) via matrix multiplication,
\(\bigO{n^3}\). Given this, we consider the Barab\'asi-Albert Scale-Free model
network (SF) as our worst-case scenario, as its degree distribution follows a
fat-tailed power-law distribution, meaning high degree nodes are likely to
occur \citep{Barabasi_1999_emergence}. Additionally, high degree nodes are also
likely to be connected to other high degree nodes. Conversely, we consider
Erd\H{o}s-R\'enyi (ER) random graphs as our best-case scenario, as the degree
distribution of ER networks is Poissonian, so the degree of nodes in the
network is likely to be more evenly distributed and high degree nodes are not
biased to connect to other high degree nodes \citep{Erdos_1960_random}. To
quantify the impact of the size of the network on the running time of our
counter, we generate instances of SF and ER networks with 20,000
nodes (which is roughly the number of nodes of the largest real networks we
consider, see Appendix Table \ref{tab:net_stats}), with the number of edges
ranging between 10,000 and 2000,000 edges. At 20,000 nodes and 2,000,000, our
synthetic networks have a density of 2.5\%, which is roughly the density of the
densest real networks we consider, See Appendix Table \ref{tab:net_stats}.  For
each setting, we generate 10 instances of each
network.  
The computations are performed single threaded and serially on an Intel Xeon
E5-2124 v6 CPU at 3.30GHz and 128GB of RAM. We present our results in Fig.
\ref{fig:times}. 

We observe that as expected, our counter is slower on SF networks than on ER.
Of course, as the density of the generated networks increases, the number of
triangles and cliques goes up more rapidly in the SF networks than in the ER
networks, which explains the widening gap in running times.  However, even in
this worst case scenario, where we consider a scale-free topology, with 20,000
nodes (larger than any of our real networks) and a density of 2.5\% (denser
than any of our real networks), our counter is able to compute all orbit
adjacencies in less than nine minutes. We conclude that GRADCO can compute
compute all orbit adjacencies in a reasonable time frame for networks up to
20,000 nodes.

\FloatBarrier

\end{document}